\documentclass[12pt,titlepage]{article}

\usepackage[margin=1in]{geometry}
\usepackage{microtype}
\usepackage{setspace}
\usepackage{enumitem}

\usepackage{amsmath,amssymb,amsthm,bm}
\numberwithin{equation}{section}

\usepackage{graphicx}
\usepackage{float}
\graphicspath{{figures_case/}} % Ensure your image folder matches this path
\usepackage{multirow}
\usepackage{booktabs}
\usepackage{threeparttable}
\usepackage{caption}
\usepackage[natbibapa]{apacite}

\usepackage{csquotes}
\usepackage{siunitx}
\usepackage{bm}

\PassOptionsToPackage{authoryear,round,semicolon,sort,longnamesfirst}{natbib}
\PassOptionsToPackage{hyphens}{url}
\usepackage{natbib}

\usepackage{xcolor} % Required for defining colors
\usepackage{hyperref}
\hypersetup{
  colorlinks=true,      % Color the text instead of boxes
  allcolors=blue,       % Set all links (refs, cites, urls) to blue
  pdftitle={A Parameterization-Invariant DIC},
  pdfauthor={Xingyao Xiao and Sophia Rabe-Hesketh},
  pdfkeywords={model comparison, information criteria, DIC, WAIC, LOO-CV}
}

\usepackage{authblk}

\usepackage[framemethod=TikZ]{mdframed}
\usepackage{listings}
\newenvironment{myenv}[1]
  {\mdfsetup{
    frametitle={\colorbox{white}{\space#1\space}},
    innertopmargin=0pt,
    frametitleaboveskip=-\ht\strutbox,
    frametitlealignment=\center,
    linewidth=1pt,
    roundcorner=10pt
    }
  \begin{mdframed}%
  }
  {\end{mdframed}}

\definecolor{darkolivegreen}{rgb}{0.33, 0.42, 0.18}

\usepackage{xspace}
\newcommand{\DIC}{\ensuremath{\mbox{DIC}}\xspace}
\newcommand{\WAIC}{\ensuremath{\mbox{WAIC}}\xspace}
\newcommand{\LOO}{\ensuremath{\mbox{LOO}}\xspace}

\newcommand{\DICpV}{\ensuremath{\mbox{DIC}_{\mbox{\tiny{}i}}}\xspace}
\newcommand{\DICpVtwo}{\ensuremath{\mbox{DIC}_{\mbox{\tiny{}p}}}\xspace}

\newcommand{\Rhatm}{\widehat{R}_{\mbox{\tiny{}MAX}}}

\newcommand{\pWAIC}{p_{\mbox{\tiny{}WAIC}}}
\newcommand{\pLOO}{p_{\mbox{\tiny{}LOO}}}

\newcommand{\pDIC}{p_{\mbox{\tiny{}DIC}}}
\newcommand{\pV}{p_{\mbox{\tiny{}V}}}

\newcommand{\Ep}[1]{\mathbb{E}_{\mathbf{\theta} \mid \mathbf{y}}\!\left[#1\right]}
\newcommand{\Vp}[1]{\mathrm{Var}_{\mathbf{\theta} \mid \mathbf{y}}\!\left[#1\right]}

\newcommand{\fm}{f_{\mathrm m}}

\newcommand{\bvec}[1]{\bm{#1}}
\newcommand{\keywords}[1]{\par\smallskip\noindent\textbf{Keywords: }#1\par}

\title{A Parameterization-Invariant DIC}
\author[1]{Xingyao Xiao}
\author[2]{Sophia Rabe-Hesketh}
\affil[1]{Stanford University}
\affil[2]{University of California, Berkeley}
\date{\today}

\begin{document}
\maketitle

\begin{abstract}
The classic Deviance Information Criterion (DIC) is not invariant to reparameterization and can have a negative and unstable effective number of parameters. The reason for the
effective number of parameters being negative is actually that the plug-in deviance becomes excessively large when the posterior means of the model parameter differ dramatically from the maximum likelihood estimates.  In latent variable models, the
cause can be identifiability issues that lead to meaningless and unstable plug-in estimates. Specifically, nonidentifiability means that distinct parameter points can have the same likelihood and switching between such points within or between MCMC chains produces unstable and meaningless posterior means.  To address this issue, we propose a plug-in-free, parameterization-invariant version of the DIC, denoted \DICpV, and show that it is asymptotically equivalent to the Watanabe-Akaike Information Criterion (WAIC). Simulations  demonstrate that \DICpV aligns with \WAIC in factor analysis and growth mixture models where the classic DIC breaks down. These results suggest that \DICpV is a useful, computationally efficient alternative to the DIC  when the \WAIC is not applicable or not available.
\keywords{identifiability, information criteria, latent variables, WAIC, LOO-CV, growth mixture models, factor analysis}
\end{abstract}

%==========================================================================
% SECTION 1: INTRODUCTION
%==========================================================================
\section{Introduction}
\label{sec:intro}

The Deviance Information Criterion \citep[DIC;][]{Spiegelhalter2002} is a popular  Bayesian information criterion in applied research, largely due to its wide availability in software such as \texttt{OpenBUGS} \citep{OpenBUGS2010}, \texttt{JAGS} \citep{Plummer2017}, and \texttt{Mplus} \citep{muthen2010bayesian}. Unfortunately, the standard version of the DIC proposed by \citet{Spiegelhalter2002} suffers from known problems, such as an unstable and negative penalty term \citep[e.g.][]{Spiegelhalter2002, plummer:2008}.  Partly for this reason, the Watanabe-Akaike Information Criterion \citep[\WAIC;][]{Watanabe2010} and leave-one-out cross-validation \citep[LOO-CV;][]{vehtari2017}, have been promoted, but they have not yet been widely adopted in Bayesian software and are not applicable for all models.

An advantage of the \WAIC and LOO-CV is that they are based on a fully Bayesian predictive distribution instead of the plug-in deviance used in the DIC. However, the \WAIC and LOO-CV require the likelihood to be factorizable into independent contributions, typically from individual units. In latent variable models, such factorization is achievable by conditioning on the latent variables. However, with this approach, the information criteria evaluate out-of-sample predictive performance for new units belonging to the \textit{in-sample} clusters. When predictive performance for out-of-sample clusters is required, the likelihood must be defined marginally over the latent variables \citep{Merkle2019}. For many models, the marginal likelihood can be factorized into contributions from a set of clusters. However, for some latent variable models, no factorization is available. Examples are spatial models where there is dependence among latent variables for different clusters, representing spatial regions, and models where the clusters associated with different latent variables are not nested, such as item response models with latent variables for subjects and items. Another problem with the \WAIC and LOO-CV is that they are computationally heavy, necessitating the storage of log-likelihood contributions from all units (or clusters) for all MCMC samples.

Latent variable models, such as finite mixtures models and factor analysis, often yield a negative DIC penalty term \citep[e.g.,][]{celeux:2006, plummer:2008, gelman2014understanding}. A negative penalty is incompatible with its interpretation as the ``optimism'' in approximating out-of-sample prediction performance by in-sample performance and contradicts its interpretation as twice the ``effective number of parameters,'' $2\pDIC$, and renders the DIC meaningless. Because the DIC penalty is the posterior mean of the deviance minus the plug-in deviance, i.e., the deviance evaluated at the posterior means, a negative penalty will occur when the plug-in deviance is excessively large. Such large plug-in deviances are  often the result of MCMC draws switching between alternative but equivalent parameterizations of the model. Consider, for example, specifying a two-component finite mixture model when the likelihood function supports a degenerate one-component model. The degeneracy can occur in two ways: either one component weight is zero or the parameters of the two component densities are identical. If posterior draws are sometimes close to one parameterization and sometimes to the other, the posterior means become meaningless, and plugging them into the deviance produces the offending large value. \citet{Xiao_Rabe-Hesketh_Skrondal_2025} call this lack of identifiability \textit{degenerate nonidentifiability}. Similarly to labeling nonidentifiability \citep{redner:84}, which leads to label switching, degenerate nonidentifiability leads to parameterization switching. Both phenomena can also be described as multimodality of the likelihood function and hence of the posterior. Because the proportion of MCMC draws near one parameterization versus the other tends to vary between estimations, the plug-in deviance and penalty term are also unstable.

Making use of an approximation of the effective number of parameters proposed by  \citet[][p.~173]{gelman2014understanding}, we define
\DICpV, where ``i'' stands for (parameterization) invariant, a criterion that does not require the plug-in deviance.
\DICpV does not exhibit instability and negative penalty terms, does not require the likelihood to be factorizable, can be computed straightforwardly from deviances evaluated at the MCMC draws, and is asymptotically equivalent (as well as numerically similar in our simulations) to WAIC and LOO-CV when all metrics are applicable. The structure of the paper is as follows. In Section~\ref{sec:models_id} we introduce two kinds of latent variable models, factor analysis and growth mixture models (GMMs), present the marginal likelihoods, and discuss identifiability issues. In Section~\ref{sec:criteria_review} we briefly review Bayesian predictive information criteria based on marginal likelihoods and discuss the problem of unstable DIC and negative DIC penalty terms. Section~\ref{sec:proposed} introduces \DICpV and shows that it is asymptotically equivalent to \WAIC. Section~\ref{sec:simulations} presents simulations for factor analysis and GMMs, and Section~\ref{sec:discussion} concludes with a discussion. The appendix provides an R function for computing \DICpV and a link to the corresponding github and interactive demo.

%==========================================================================
% SECTION 2: LATENT VARIABLE MODELS AND BAYESIAN IDENTIFIABILITY
%==========================================================================
\section{Latent Variable Models and Bayesian Identifiability}
\label{sec:models_id}

\subsection{Latent variable models}
\label{subsec:general_framework}

We consider models that include continuous, sometimes multivariate, latent variables $\bvec{\eta}_j$ and/or discrete, typically univariate, latent variables $\xi_j$
associated with clusters $j$ that comprise $n_j$ units with corresponding $n_j$-dimensional response vectors $\bvec{y}_j$, $j=1,\ldots,J$. One example is an item response model for students $j$ whose ability $\eta_j$ is measured by responses to $n_j$ test questions.  Another example is a multilevel model for students nested in schools, where $\bvec{\eta}_j$ are varying intercepts and varying slopes of student-level covariates, the units are now students and the clusters are schools.
In latent class or finite mixture models, $\xi_j$ is a discrete variable denoting a subject's membership in a class or mixture component, and $\bvec{y}_j$ is a response vector for $n_j=n$ different variables. When only one form of clustering is present, the posterior is proportional to
\begin{equation}
f_1(\bvec{y}_{j}|\bvec{\eta}_j,\xi_j,\bvec{\theta}_1)
f_2(\bvec{\eta}_j,\xi_j|\bvec{\theta}_2)p_2(\bvec{\theta}_1)p_3(\bvec{\theta}_2).
\label{eq:hierBayes}\end{equation}
At Stage 1, we have the conditional distribution $f_1(\bvec{y}_{j}|\bvec{\eta}_j,\xi_j,\bvec{\theta}_1)$ of the responses given the latent variables and stage-1 parameters $\bvec{\theta}_1$. This term is often viewed as the likelihood, and we refer to it as the \textit{conditional likelihood}. At Stage 2 are the corresponding priors   $f_2(\bvec{\eta}_j,\xi_j|\bvec{\theta}_2)$ and $p_2(\bvec{\theta}_1)$. What makes the model hierarchical is that the priors of the latent variables depend on free hyperparameters $\bvec{\theta}_2$ that have hyperpriors $p_3(\bvec{\theta}_2)$ at stage 3. Treating  $f_1(\bvec{y}_{j}|\bvec{\eta}_j,\xi_j,\bvec{\theta}_1)$  as the likelihood and the last three terms as priors produces a posterior predictive distribution that can predict new responses for the clusters included in the data. However, latent variable models are usually specified to make inferences regarding the \textit{population} of clusters and hence the relevant likelihood is $f_1(\bvec{y}_{j}|\bvec{\eta}_j,\xi_j,\bvec{\theta}_1)
f_2(\bvec{\eta}_j,\xi_j|\bvec{\theta}_2)$ marginalized over the latent variables, with the last two terms in (\ref{eq:hierBayes}) representing the priors. Such ``marginal'' (over the latent variables) likelihoods produce  mixed predictive distributions \citep{gelman:96}. Following \citet{Merkle2019}, we will use mixed predictive distributions as the basis for the predictive information criteria discussed in Section 3.
Letting $\bvec{y}$ be the response vector across all clusters and  $\bvec{\theta}^{\prime}=(\bvec{\theta}_1^{\prime},\bvec{\theta}_2^{\prime})$ the parameter vector of the marginal likelihood, we denote the marginal likelihood for the dataset $\fm(\bvec{y} \mid
\bvec{\theta})$. The corresponding  \emph{marginal deviance} is
\begin{equation}
D(\bvec{\theta}) \;=\; -2 \log \fm(\bvec{y} \mid
\bvec{\theta}).
\label{eq:marginal_deviance}
\end{equation}

For concreteness, we now present two types of latent variable models that will be used in the simulation studies in Section 5, giving the form of both conditional and marginal likelihoods.

\subsection{Factor Analysis}
\label{subsec:model_fa}

We consider a confirmatory factor analysis model with $n$ continuous indicators. For individual $j$, the response vector $\bvec{y}_j = (y_{1j}, \dots, y_{n})^{\prime}$ depends on a single latent factor $\eta_j \sim \mathcal{N}(0,1)$. The conditional likelihood is
\begin{equation}
\bvec{y}_j \mid \eta_j, \bvec{\theta}_1 \sim \mathcal{N}_n\big(\bvec{\mu}
+ \bvec{\lambda}\eta_j, \bvec{\Sigma}\big),
\label{eq:fa_conditional}
\end{equation}
where $\bvec{\mu}$ is the $n$-dimensional vector of means or intercepts, $\bvec{\lambda}$ is the $n$-dimensional
vector of factor loadings, and $\bvec{\Sigma} = \mathrm{diag}(\sigma_1^2,
\dots, \sigma_n^2)$ contains the unique variances.
Here $\bvec{\theta}=\bvec{\theta}_1$ comprises $\bvec{\mu}$, $\bvec{\lambda}$, and the diagonal elements of $\bvec{\Sigma}$, and there are no hyperparameters $\bvec{\theta}_2$ because the factor variance has been constrained to 1.

Integrating out $\eta_j$
yields the marginal likelihood
\begin{equation}
\bvec{y}_j \mid \bvec{\theta} \sim \mathcal{N}_n\big(\bvec{\mu},
\bvec{V}(\bvec{\theta})\big), \quad \text{with }
\bvec{V}(\bvec{\theta}) = \bvec{\lambda}\bvec{\lambda}^{\prime}
+ \bvec{\Sigma}.
\label{eq:fa_marginal}
\end{equation}

We assign  priors
$\mu_i \sim \mathcal{N}(0, 1)$ and
 $\sigma_i = 0.10 + \exp(\tau_i)$ with
$\tau_i \sim \mathcal{N}(\log 0.6,\; 0.25^2)$ for $i=1,\ldots,n$.

\subsection{Growth Mixture Models}
\label{subsec:model_gmm}

A popular approach for analyzing heterogeneity in developmental trajectories is the growth mixture model (GMM), a finite mixture of linear mixed models for growth \citep[e.g.,][]{muthen:99,muthen:2000,muthen:2002}. For example, \citet{Xiao_Rabe-Hesketh_Skrondal_2025} used GMMs to model heterogeneous reading skill trajectories in the National Longitudinal Survey of Youth, and we will consider their model here. Other applications include criminal trajectory analysis \citep{Kreuter2008AnalyzingCT} and child achievement trajectories \citep{Pianta:2008}. Each of $K$ subpopulation or latent class has its own mean growth trajectory, quadratic in time here, with intercepts and slopes of time varying between individuals. The covariance matrices of the  varying intercepts and slopes are class specific. For individual $j$ who belongs to class $k$, denoted $\xi_j=k$, the conditional likelihood of the vector of responses $\bvec{y}_j$ across $n_j$ measurement occasions can be written as
\[
\bvec{y}_j| \xi_j=k, \bvec{\eta}_j, \bvec{\theta}_1 \ \sim \ \mathcal{N}_{n_j}\big( \bvec{X}_j \bvec{\beta}^{(k)} + \bvec{Z}_j \bvec{\eta}_j,\,
\sigma_e^2 \bvec{I}_{n_j}\big),
\]
where $\bvec{X}_j$ and $\bvec{Z}_j$ are $n_j\times 3$
and $n_j\times 2$ design matrices, with first columns
equal to 1, second columns equal to the times $t_{ij}$
associated with the measurement occasions, and third
column of $\bvec{X}_j$ equal to $t_{ij}^2$. The parameters are $\bvec{\theta}_1^{\prime}=(\bvec{\beta}^{(1)\prime}, \ldots, \bvec{\beta}^{(K)\prime}, \sigma_e^2)$, where
$\bvec{\beta}^{(k)} = (\beta_0^{(k)}, \beta_1^{(k)},
\beta_2^{(k)})^{\prime}$ are the class-specific fixed intercept,
slope of $t_{ij}$, and slope of the quadratic term $t_{ij}^2$. The latent variables
$\bvec{\eta}_j = (\eta_{0j}, \eta_{1j})^{\prime}$ are a varying
intercept and a varying slope of $t_{ij}$.
Their distributions are specified as
\[
\bvec{\eta}_j | \xi_j=k,\bvec{\theta}_2 \ \sim\  \mathcal{N}_{2}(\bvec{0}, \bvec{\Psi}^{(k)}), \quad
p(\xi_j=k|\bvec{\theta}_2)=\pi^{(k)},
\]
with $\bvec{\theta}_2$ comprising the unique elements of the $K$ class-specific covariance matrices $\bvec{\Psi}^{(k)}$,   and the $K$ component weights or class probabilities $\pi^{(k)}$, with $\sum_{k=1}^K \pi^{(k)}=1$.

The class-specific marginal likelihood, integrated over $\bvec{\eta}_j$, is a multivariate normal density, $\phi_{n_j}\big(\bvec{y}_j \mid \bvec{X}_j \bvec{\beta}^{(k)}, \bvec{V}_j^{(k)}\big)$, with covariance
matrix $\bvec{V}_j^{(k)} = \bvec{Z}_j \bvec{\Psi}^{(k)} \bvec{Z}_j^{\prime} + \sigma_e^2 \bvec{I}_{n_j}$. Summing over the $K$ latent classes, the  marginal likelihood becomes
\begin{equation}
f(\bvec{y}_j \mid \bvec{\theta}) = \sum_{k=1}^K \pi^{(k)}\,\phi_{n_j}\big(\bvec{y}_j \mid \bvec{X}_j \bvec{\beta}^{(k)}, \bvec{V}_j^{(k)}\big).
\label{eq:gmm_marginal}
\end{equation}
We specify normal priors for the fixed intercepts and slopes, half-normal priors for $\sigma_e$ and for the standard deviations in
$\bvec{\Psi}^{(k)}$, $\text{LKJ}(\nu)$ priors \citep{Lewandowski:2009} for the
corresponding correlation matrices, and a
$\mathrm{Dirichlet}(\alpha\bvec{1}_K)$ prior
for the mixing proportions, with concentration parameter $\alpha$.

\subsection{Identifiability and parameterization switching}
\label{subsec:identifiability}
As discussed in \citet{Xiao_Rabe-Hesketh_Skrondal_2025}, Bayesian identifiability of latent variable models corresponds to likelihood identifiability if the marginal likelihood is considered and if the prior of  $\bvec{\theta}$ has support on the full parameter space. Investigation of likelihood identifiability involves asking whether there are observationally equivalent parameter points (or values), so that for all possible data, the likelihood at the different parameter values is identical. For example, for a simple two-component mixture model with component weights $\pi^{(k)}$ and class-specific parameters $\bvec{\beta}^{(k)}$, the parameter point  ($\pi^{(1)}=\delta$, $\pi^{(2)}=1-\delta$, $\bvec{\beta}^{(1)}=\bvec{\gamma}_1$, $\bvec{\beta}^{(2)}=\bvec{\gamma}_2$) is observationally equivalent to the parameter point
 ($\pi^{(1)}=1-\delta$, $\pi^{(2)}=\delta$, $\bvec{\beta}^{(1)}=\bvec{\gamma}_2$, $\bvec{\beta}^{(2)}=\bvec{\gamma}_1$), a phenomenon known as \textit{labeling nonidentifiability} \citep{redner:84}. When such sets of observationally equivalent parameter points exist, the likelihood is not globally identified.

 In maximum likelihood estimation, global identifiability is not necessary for achieving convergence if the  likelihood is \textit{locally} identified at all parameter points (or even just at the maximum likelihood estimate (MLE), called empirical local identification).
 The likelihood is locally identified at a parameter point if there exists an open neighborhood at the parameter point in which there is no other point that is observationally equivalent. Violations of local identifiability are usually due to the likelihood function being flat in some direction which causes convergence problems in maximum likelihood estimation. For locally but not globally identified models, maximum likelihood procedures usually converge to one of the observationally equivalent parameter points that maximize the likelihood.

 If local identifiability holds, global non-identifiability can also be described as multimodality of the likelihood function and hence of the posterior.
 Unlike maximum likelihood estimation, standard Bayesian estimation via MCMC sampling is adversely affected by multimodality because the samples may switch between modes \citep[e.g.][]{stacking}. In the case of labeling nonidentifiability, this phenomenon is well known as label switching \citep{diebolt:94}. If such switching occurs, posterior summaries, such as posterior means, will no longer be meaningful.

 We now describe two other kinds of switching that occur, namely sign switching in factor analysis, due to reflection invariance, and parameterization switching in GMMs, due to degenerate nonidentifiability. These kinds of switching, as well as label switching, are examples of \textit{parameterization switching} because switching occurs between observationally equivalent
parameter points, and these datapoints can be described as different parameterizations, all yielding the same likelihood.

 The marginal multivariate normal likelihood of the factor analysis model is invariant to reflection of the factor loadings $\bvec{\lambda}$ because the covariance matrix
 $\bvec{V}(\bvec{\theta}) = \bvec{\lambda}\bvec{\lambda}^{\prime}
+ \bvec{\Sigma}$ depends on $\bvec{\lambda}$ only through
$\bvec{\lambda}\bvec{\lambda}^{\prime}$. Consequently, sign switching will occur in Bayesian estimation \citep[e.g.,][]{Merkle:2021}.
Although researchers often attempt to resolve this problem by constraining a
loading to be positive (e.g., $\lambda_1 > 0$), such constraints create
artificial boundaries in the parameter space and fail to resolve multimodal mixing issues in the remaining
parameters \citep{erosheva2017_FA, papastamoulis2022_FA}.

As demonstrated by \citet{Xiao_Rabe-Hesketh_Skrondal_2025},
 \emph{degenerate nonidentifiability} is a problem for Bayesian estimation of GMMs and other finite mixture models. When there are $K$ classes or mixture components, there are several observationally equivalent parameter points that are degenerate in the sense that  there is a corresponding parameter point for a finite mixture model with fewer than $K$ components that yields the same likelihood. For example, consider $K=3$ where the two-component parameter point of interest is ($\pi^{(1)}=\delta$, $\pi^{(2)}=1-\delta$, $\bvec{\beta}^{(1)}=\bvec{\gamma}_1$, $\bvec{\beta}^{(2)}=\bvec{\gamma}_2$). Apart from label switching, there are three ways of generating equivalent degenerate parameter points for the $K=3$ model that correspond to this point: (1) setting one component weight to zero (empty class), e.g., $\pi^{(3)}=0$; (2) setting one pair of component-specific parameters equal (merged classes), e.g., $\bvec{\beta}^{(3)}=\bvec{\beta}^{(1)}=\bvec{\gamma}_1$; and (3) setting another pair equal, e.g., $\bvec{\beta}^{(3)}=\bvec{\beta}^{(2)}=\bvec{\gamma}_2$. There are even more ways of generating parameter points that correspond to a parameter point with $K=1$. As the number of components of the specified model increases, the number of equivalent degenerate parameter points increases rapidly.
 If the Markov chain visits any of these degenerate points, it can switch to any of the other equivalent degenerate points.

Interestingly, whether empty classes or merged classes are more likely to occur can be influenced by the choice of concentration parameter
for the Dirichlet prior.
As shown by \citet{rousseau:2011} for a general class
of finite mixture models, overfitting (i.e., specifying $K$ greater than the true number of components) asymptotically results
in classes being empty if the concentration parameter satisfies $\alpha < d/2$ and classes merging
if $\alpha > d/2$, where $d$ is the number of class-specific
parameters. The GMM described in this section has $d=6$ class-specific parameters ($\beta_0^{(k)}$,
$\beta_1^{(k)}$, $\beta_2^{(k)}$, and three unique elements of $\bvec{\Sigma}^{(k)}$).

Parameterization switching causes a problem for parameter estimation because the posterior mean will lie somewhere between
posterior modes and will essentially be meaningless. If inference regarding parameters is of interest, this problem could be resolved by postprocessing, as is often done for label switching \citep[e.g.][]{stephens:2000}
and sign switching \citep[e.g.][]{erosheva2017_FA}. In principle, such procedures could be developed for switching due to degenerate
non-identifiability. Such post-processing is not necessary for inferences regarding functions of the parameters that
are  invariant to the reparameterizations, such as the  likelihood or deviance. We will see in Section 3 that some Bayesian information criteria are invariant to reparameterization whereas the classic DIC is not because it relies on point estimates of the parameters.

%==========================================================================
% SECTION 3: BAYESIAN PREDICTIVE INFORMATION CRITERIA (Review)
%==========================================================================
\section{Bayesian Predictive Information Criteria}
\label{sec:criteria_review}

Predictive information criteria aim to estimate the
expected out-of-sample predictive accuracy of a model.
The logic behind these criteria is shared with the classic AIC
\citep{Akaike1973}. Namely, using the same data to fit the model
and evaluate the fit produces an optimistic
assessment of predictive accuracy. To correct for this
optimism, each criterion adds a penalty term that
approximates the expected difference between
out-of-sample fit and in-sample fit.  The criteria differ in how they
define fit and how they estimate the penalty.

\subsection{Classic DIC and Its Failure}
\label{subsec:dic_classic}

The DIC \citep{Spiegelhalter2002} adapts the logic of
AIC to the Bayesian setting. Whereas
the AIC uses the deviance at the MLEs as the definition of fit, the DIC uses the deviance at the Bayesian
point estimate, often referred to as the plug-in deviance. The  point estimate
$\tilde{\bvec{\theta}}$ is typically an MCMC estimate of the posterior mean
$\Ep{\bvec{\theta}}$.
The penalty term for optimism due to evaluating the deviance at the in-sample data is twice
the effective number of parameters $\pDIC$ defined as
\begin{equation}
\pDIC \;=\; \Ep{D(\bvec{\theta})} - D(\tilde{\bvec{\theta}}),
\label{eq:pdic}
\end{equation}
and the DIC can then be written as
\begin{equation}
\DIC \;=\; D(\tilde{\bvec{\theta}}) + 2\pDIC.
\label{eq:dic_classic}
\end{equation}
The first term in (\ref{eq:pdic}) is approximated by the sample mean of the marginal deviance evaluated at $S$ draws of $\bvec{\theta}$ from its posterior distribution. This marginal form of the DIC was discussed by \citet{Spiegelhalter2002} for when $\bvec{\theta}$ is in focus (and not the latent variables), was called $\mbox{DIC}_1$ by \citet{celeux:2006}, and was recommended as the best choice for latent variable models by \citet{Merkle2019}.

Although $\pDIC$ is non-negative for log-concave posterior densities, it is often negative in practice \citep[e.g.,][]{Spiegelhalter2002, celeux:2006, plummer:2008}.
As discussed in Section~\ref{subsec:identifiability},
switching between equivalent parameterizations renders $\tilde{\bvec{\theta}}$
meaningless, leading to extremely large plug-in deviances
$D(\tilde{\bvec{\theta}})$ and, consequently, negative values for $\pDIC$ \citep[see also][]{celeux:2006,plummer:2008}.
Furthermore, $\pDIC$ tends to vary greatly between MCMC runs because the proportion of posterior samples of $\bvec{\theta}$
near each of the posterior modes varies between runs, leading to highly variable  posterior means and hence plug-in deviances.
Negative $\pDIC$ can occur for other reasons whenever the posterior mean of $\bvec{\theta}$
is far from its MLE, as plugging the MLE into the deviance would minimize the deviance. Large differences between posterior means and MLEs can be due to a ``substantial conflict between prior and data'' \citep{Spiegelhalter2002}, or due to the posterior mean being far from the posterior mode \citep{gelman2014understanding}, for example when the posterior distribution is extremely skewed.

\subsection{\WAIC\ and LOO-CV}
\label{subsec:waic_loo}

Let $\ell_j(\bvec{\theta}) = \log f(\bvec{y}_j \mid
\bvec{\theta})$ denote the pointwise marginal
log-likelihood contribution for cluster $j$ so that the log-likelihood for the data is $\sum_{j=1}^J \ell_j(\bvec{\theta})$.  Then the marginal version of the \WAIC\ \citep{Watanabe2010} is defined as \citep{Merkle2019}
\begin{equation}
\WAIC \;=\; -2\sum_{j=1}^J \log \Ep{f(\bvec{y}_j \mid
\bvec{\theta})} \;+\; 2\sum_{j=1}^J \Vp{\ell_j (\bvec{\theta})}.
\label{eq:waic}
          \end{equation}
The posterior expectations $\Ep{f(\bvec{y}_j \mid
\bvec{\theta})}$ of the likelihood contributions are also point predictive densities, and the first term of the \WAIC\ is often
referred to as the log pointwise predictive density, lppd. Because the marginal likelihood is used, $\Ep{f(\bvec{y}_j \mid
\bvec{\theta})}$ are actually \textit{mixed} predictive densities, as defined by \citet{gelman:96}, rather than
posterior predictive densities.
As usual, the posterior expectation is approximated by the average over $S$ posterior samples. Analogously, the posterior variances for the effective number of parameters $\pWAIC = \sum_{j=1}^J
\Vp{\ell_j(\bvec{\theta})}$ are approximated by the sample variances of the log-likelihood contributions across the $S$ posterior samples.

WAIC evaluates the lppd for the in-sample data and adds a penalty term to approximate
the expected out-of-sample lppd. To avoid the need for a penalty term, leave-one-out cross-validation could be used. LOO-CV approximates the leave-one-out version of the lppd by using
Pareto-smoothed importance weights \citep{vehtari2017}.

As pointed out by \citet[][p. 173, 174]{gelman:2014}, the \WAIC\
and LOO-CV are more fully Bayesian than the DIC because they are based on posterior (or mixed) predictive distributions
instead of conditioning on a point estimate. Another advantage that is particularly important for latent
variable models is that the  \WAIC\ and  LOO-CV are parameterization invariant and hence  robust to the multimodality discussed
in Section~\ref{subsec:identifiability}. However, a disadvantage is that these criteria require
storing the full $J \times S$ matrix of pointwise
log-likelihood contributions, whereas the DIC requires only the deviance of the entire dataset at the parameter draws. Another disadvantage of the
\WAIC\
and LOO-CV is that they require the likelihood to be factorizable.

\subsection{Alternative DIC Formulations}
\label{subsec:alternatives}

Several modifications to the classic DIC have been
proposed to address the problem of negative effective number of parameters, avoid the plug-in deviance, and/or extend the
DIC definition to latent variable models.
\citet{gelman2014understanding} propose replacing
$\pDIC$ with half the posterior variance of the deviance,
\begin{equation}
\pV \;=\; \tfrac{1}{2}\,\Vp{D(\bvec{\theta})},
\label{eq:pV}
\end{equation}
which is always non-negative and invariant to
reparameterization. However, they retain the plug-in deviance in the first term, defining
\begin{equation}
\DICpVtwo \;=\; D(\tilde{\bvec{\theta}}) + 2\pV.
\label{eq:dic_pv2}
\end{equation}
In practice, $\pV$ is approximated by the sample variance of $D(\bvec{\theta})$ across $S$ posterior samples.
Although  $\pV$ ``generally turns out to be remarkably robust and accurate'' \citep{Spiegelhalter2014}, the
first term in $\DICpVtwo$ can be grossly inflated, for example when
$\tilde{\bvec{\theta}}$ falls between posterior modes, as discussed in Section~\ref{subsec:identifiability}.

\citet{celeux:2006} defined eight versions of the DIC for latent variable models, the first three of which involve the marginal likelihood. $\mbox{DIC}_1$ is the version we defined
in Section 3.1 and $\mbox{DIC}_2$ is the same except that it uses the posterior mode for $\tilde{\bvec{\theta}}$.
$\mbox{DIC}_3$ was proposed by \citet{Richardson:2002} in the context of finite mixtures of normal densities. Instead of plugging the posterior expectation of the parameters $\bvec{\theta}$ into the plug-in deviance, the posterior expectation $\Ep{f(\bvec{y}_j|\theta)}$  of the marginal density is plugged in, which is invariant to reparameterization. Both  \citet{Richardson:2002} and \citet{celeux:2006}  found that the corresponding effective number of parameters for $\mbox{DIC}_3$ stabilize with increasing number of mixture components when the additional components no longer appreciably change  the shape of the marginal density. While this lack of impact of the additional parameters on the marginal density should be reflected by a large increase in the DIC, the penalty inadequately represents the increasing model complexity.
\citet{Du2024} compared the performance of $\mathrm{DIC}_1$ with what \citet{celeux:2006} call $\mathrm{DIC}_3$ (and they called $\mathrm{DIC}_2$). They studied these criteria for multilevel/hierarchical linear models (with multiply imputed missing data) for choosing among models that included the data-generating model and models that had one more or one fewer fixed coefficients or one more or one fewer random coefficients than the data-generating model. They found that the marginal versions performed better than the conditional versions and that  $\mathrm{DIC}_3$ compared favorably with $\mathrm{DIC}_1$ and WAIC.

Celeux et al.'s $\mbox{DIC}_4$ to $\mbox{DIC}_6$ define the expected deviance in the DIC as the posterior expectation of $D(\bvec{\theta},\bvec{\xi})$ over the joint posterior distribution of $\bvec{\theta}$ and $\bvec{\xi}$,
with three different versions of the plug-in deviance. (They do not distinguish between discrete and continuous latent variables, so $\bvec{\xi}$ could be replaced by both  $\bvec{\xi}$ and $\bvec{\eta}$.) Perhaps the most persuasive of these versions of the DIC is $\mbox{DIC}_4$ which  uses the posterior expectation of $\bvec{\theta}$ over the joint posterior of $\bvec{\theta}$ and $\bvec{\xi}$ for the plug-in deviance.
While $\mbox{DIC}_4$-$\mbox{DIC}_6$ can be computed relatively easily when the $\bvec{\xi}$ is sampled from its posterior, it is not clear how to compute them when that is not the case, as in \texttt{Stan} when $\bvec{\xi}$ includes discrete latent variables.
Finally, $\mbox{DIC}_7$ and $\mbox{DIC}_8$ treat the latent variables as model parameters and hence use conditional likelihoods, which are not considered in this paper. \citet{Li2020} show that $\mbox{DIC}_7$
is asymptotically biased due to the incidental parameter problem.

\citet{plummer:2008} defines a different target quantity for the DIC, namely  the \textit{posterior mean} deviance for out-of-sample data, instead of relying on the plug-in deviance. One advantage for this target quantity, pointed out by Plummer,  is that it is not sensitive to reparameterization and is hence ``coordinate free.'' In practice,
  the posterior expectation of the in-sample deviance is used, $\Ep{D(\bvec{\theta})}$, and a penalty term is added. Computing the penalty term requires simulations of new responses from $f(\bvec{y}|\bvec{\theta})$ for posterior draws of
$\bvec{\theta}$ from at least two parallel chains and is hence computationally complex. This version of the DIC, provided by the \texttt{JAGS} software \citep{Plummer2017}, does not rely on plug-in estimates and is invariant to reparameterization.

%==========================================================================
% SECTION 4: THE PROPOSED DIC
%==========================================================================
\section{The Proposed \DICpV}
\label{sec:proposed}

\subsection{Motivation and Definition}
\label{subsec:dicpv_def}

By plugging~\eqref{eq:pdic} into ~\eqref{eq:dic_classic}, we can write the classic \DIC  as
\begin{eqnarray*}
\DIC &=& D(\tilde{\bvec{\theta}}) + 2\{\Ep{D(\bvec{\theta})} - D(\tilde{\bvec{\theta}})\} \\
  &=&\Ep{D(\bvec{\theta})}  +  \{ \Ep{D(\bvec{\theta})} - D(\tilde{\bvec{\theta}})\}\\
  &=&\Ep{D(\bvec{\theta})}+ \pDIC.
\end{eqnarray*}
We propose replacing
$\pDIC$ with $\pV$ defined in \eqref{eq:pV} in the last row of this equation, giving
\begin{equation}
\label{eq:dicpv}
\DICpV \;=\; \Ep{D(\bvec{\theta})} \;+\; \pV.
\end{equation}
Both terms in \DICpV\ are non-negative and invariant to
parameterization.

Alternatively, we could define $\DICpV$ as the average
$(\DIC + \DICpVtwo)/2$, or, equivalently,
replace $\pDIC$ in ~\eqref{eq:dic_classic} by the average $(\pDIC + \pV)/2$, giving
\begin{eqnarray*}
\DICpV &=& \{ D(\tilde{\bvec{\theta}}) + \pDIC \} + \pV \\
&=&  \Ep{D(\bvec{\theta})}  + \pV.
\end{eqnarray*}
We also see that
\[
\DICpVtwo - \DICpV \;=\; \pV - \pDIC \;=\; \DICpV - \DIC.
\]
When $\pDIC$ is negative, $\DICpVtwo$ will be larger than
$\DICpV$ by $\pV - \pDIC$, whereas $\DIC$ will be smaller than
$\DICpV$ by the same amount. Therefore, both $\DICpVtwo$ and $\DIC$ will be equally extreme (in opposite directions) and unstable when $\pDIC$ is negative and unstable.

\subsection{Asymptotic Equivalence to \WAIC}
\label{subsec:dicpv_waic_relation}

As in Section~\ref{subsec:waic_loo}, we assume that the likelihood factorizes into pointwise contributions, where the ``points''  are clusters $j$ when the model includes cluster-specific latent variables and the likelihood is marginal over the latent variables~\citep{Merkle2019},
\[
D(\bvec{\theta}) = -2\sum_{j=1}^J\ell_j(\bvec{\theta}).
\]

To compare \WAIC and \DICpV asymptotically, we can expand the log point-predictive densities in the first term of the WAIC using the cumulant-generating function \citep[see, e.g.,][Ch.~2]{mccullagh1987tensor}
\begin{equation}
\log \Ep{f(\bvec{y}_j\mid\bvec{\theta})}
\;=\; \Ep{\ell_j(\bvec{\theta})}
\;+\; \tfrac{1}{2}\Vp{\ell_j(\bvec{\theta})}
\;+\; R_j,
\label{eq:cumulant_expansion}
\end{equation}
where $R_j$ collects higher-order posterior cumulants. In regular models with asymptotically normal posteriors, $R_j = O_p(J^{-2})$ and is negligible \citep[cf.][Ch.~10]{vanderVaart1998}.
Summing~\eqref{eq:cumulant_expansion} over $J$ clusters and multiplying by $-2$, we obtain the following expression for \WAIC
\begin{align}
\WAIC &= -2\sum_{j=1}^J \log \Ep{f(\bvec{y}_j\mid\bvec{\theta})} + 2\sum_{j=1}^J \Vp{\ell_j(\bvec{\theta})}\\
&= -2\sum_{j=1}^J \left( \Ep{\ell_j(\bvec{\theta})} + \tfrac{1}{2}\Vp{\ell_j(\bvec{\theta})} + R_j \right) + 2\sum_{j=1}^J \Vp{\ell_j(\bvec{\theta})}\nonumber \\
&= \Ep{D(\bvec{\theta})} + \sum_{j=1}^J\Vp{\ell_j(\bvec{\theta})} - 2\sum_{j=1}^J R_j.
\label{eq:fit_alignment}
\end{align}

The first term of the \WAIC is identical to the first term of the \DICpV. Asymptotically, $\sum_{j=1}^JR_j$ vanishes, so it remains to compare $\sum_{j=1}^J\Vp{\ell_j(\bvec{\theta})}$ with $\frac{1}{2}\Vp{D(\bvec{\theta})}$. Under standard regularity conditions, the Bernstein-von Mises theorem implies that the posterior distribution of $\bvec{\theta}$ converges to a multivariate normal distribution centered at the maximum likelihood estimate \citep[Ch.~10]{vanderVaart1998}. A direct consequence of this asymptotic normality is that the posterior deviance, $D(\bvec{\theta})$, converges in distribution to a $\chi^2_q$ variable shifted by the deviance at the mode \citep[p.~173]{gelman:2014},
\begin{equation}
\label{eq:chi2}
D(\bvec{y} \mid \bvec{\theta}) - D(\bvec{y} \mid \hat{\bvec{\theta}}_{\mathrm{MLE}})
\;\xrightarrow{d}\; \chi^2_q\,,
\end{equation}
where $q$ is the number of parameters. Since the variance of a $\chi^2_q$ random variable is $2q$,  this implies that $\pV = \tfrac{1}{2}\Vp{D(\bvec{\theta})}$ converges to $q$. It is also well established that $\pWAIC = \sum_{j=1}^J\Vp{\ell_j(\bvec{\theta})}$ converges to $q$  \citep{gelman2014understanding,Watanabe2010}. Because both penalties converge to the same limit and $\sum_j R_j = O_p(J^{-1})$, we see that \DICpV and $\textrm{WAIC}$ converge to the same limit.
In Sections~\ref{sec:fa} and~\ref{sec:gmm}, we demonstrate that \DICpV and \WAIC are closely aligned in practice, with $\pV \approx \pWAIC$.

%==========================================================================
% SECTION 5: SIMULATION STUDIES
%==========================================================================
\section{Simulation Studies}
\label{sec:simulations}
In this section we assess performance of the information criteria for the factor analysis model  and GMM described in Section 2.
All models are estimated using Hamiltonian Monte Carlo
(HMC) via \texttt{CmdStan} \citep{Stan2021}.

%--------------------------------------------------------------------------
% Factor Analysis
%--------------------------------------------------------------------------
\subsection{Factor analysis model}
\label{sec:fa}

\subsubsection{Simulation Design}
\label{sec:fa_design}

\paragraph{Data Generation.}
We simulated data for the one-factor model defined in Section~\ref{subsec:model_fa} with $n=6$ continuous indicators. The factor loadings were specified as $\bvec{\lambda} = c \cdot (0.9, 0.8, 0.7, 0.6, 0.5, 0.4)^{\prime}$, where $c$ scales the factor strength, and the unique variances were homogeneous, $\bvec{\Sigma} = \sigma^2 \bvec{I}_6$. We utilized a full factorial design crossing three factors:
\begin{enumerate}[label=(\roman*), nosep]
    \item \textbf{Factor Strength:} Weak ($c=0.3$), Moderate ($c=0.6$), and Strong ($c=0.9$);
    \item \textbf{Measurement Noise:} Low ($\sigma^2=0.5$) and High ($\sigma^2=1.0$);
    \item \textbf{Sample Size:} $J=400$ and $J=800$.
\end{enumerate}
This yielded 12 simulation conditions, each replicated 100 times.

\paragraph{Estimation.}
An unconstrained unidimensional factor analysis model was estimated for each dataset, with $q=18$ free parameters (6~intercepts, 6~loadings, 6~unique variances).
 Priors were specified as described in Section~\ref{subsec:model_fa}. Importantly, we assigned symmetric priors to the loadings, $\lambda_j \sim \mathcal{N}(0,1)$, so that negative and positive loadings were equally likely a priori. For each replicate dataset, 4 chains were run with 1,000 warmup draws and 1{,}000 post-warmup draws.
Starting values were generated using \texttt{CmdStan}'s
default initialization, which draws uniformly on $(-2, 2)$
in the unconstrained parameter space
\citep{CmdStanManual2024}. Because this initialization is
symmetric around zero, chains are equally likely to begin with positive
or negative loadings.

\subsubsection{Results}
\label{sec:fa_results}

\begin{figure}[htbp]
    \centering
    \includegraphics[width=0.95\linewidth]{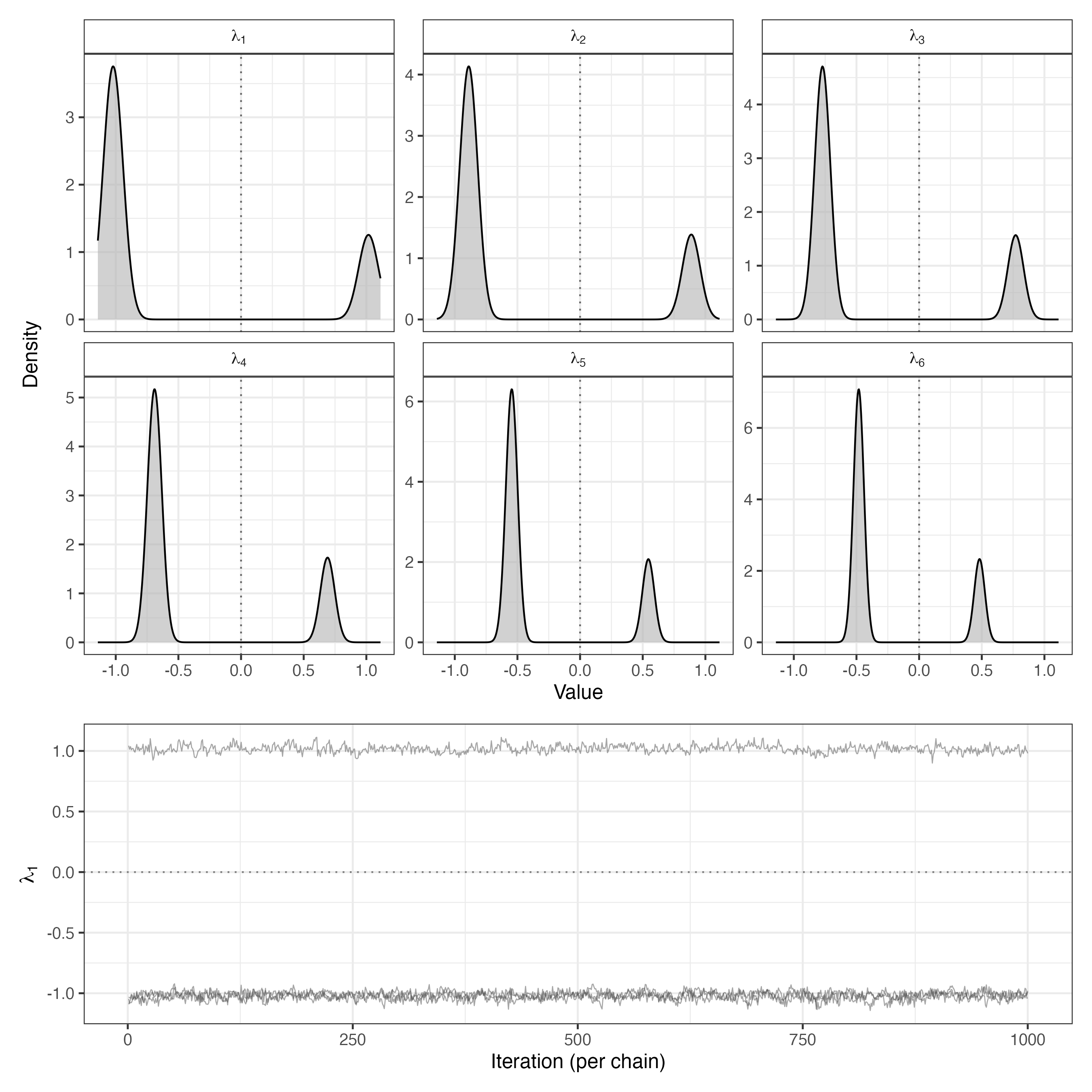}
    \caption{Marginal posterior densities of the six factor loadings, pooled across chains (top panel), and per-chain trace-plots for $\lambda_1$ (bottom panel), for one simulated dataset ($c = 0.9$, $\sigma^2 = 1.0$, $J = 800$, replicate~50).}
    \label{fig:FA_posteriors}
\end{figure}

\paragraph{Posterior behavior.}
Figure~\ref{fig:FA_posteriors} illustrates the sign switching of the factor loadings for a representative replicate
($c = 0.9$, $\sigma^2 = 1.0$, $J = 800$, replicate~50).
The posterior densities of all six loadings are bimodal
(top panel), with each chain settling into either the
positive or negative mode for the duration of the sampling
(bottom panel). In this replicate, Chain~1 converged to the
negative mode while Chains~2--4 converged to the positive
mode. For each replicate, we computed the posterior mean of each factor loading separately within each of the four chains and classified the replicate as exhibiting between-chain sign-switching if the chain-specific posterior means did not all share the same sign across chains. Comparing this classification with the sign of $\pDIC$, we found exact correspondence across all 1{,}200 replications: the 1{,}065 (89\% of 1{,}200) replicates with $\pDIC < 0$
all exhibited between-chain sign-switching, and the 135
replicates with $\pDIC > 0$ all did not. We therefore use
$\pDIC < 0$ as a diagnostic for between-chain sign-switching
throughout.

For the replicate in the figure, the rank-normalized split-$\widehat{R}$ \citep{vehtari:2021} produced by default by
 both \texttt{CmdStanR} and the
\texttt{posterior} \textsf{R} package \citep{burkner2023posterior} yielded $\widehat{R} \approx 1.5$
for all six loading parameters. While this exceeds recommended thresholds, it does not reflect
the massive between-chain variance as well as the classic
Gelman--Rubin $\widehat{R}$ \citep{Gelman:92}, available via
\texttt{posterior::rhat\_basic()}, that took on  values
between $19$ and $30$ for the same parameters. Intercepts and residual
standard deviations had both versions of $\widehat{R}$ close to $1.00$.

\begin{figure}[htbp]
    \centering
    \includegraphics[width=\linewidth]{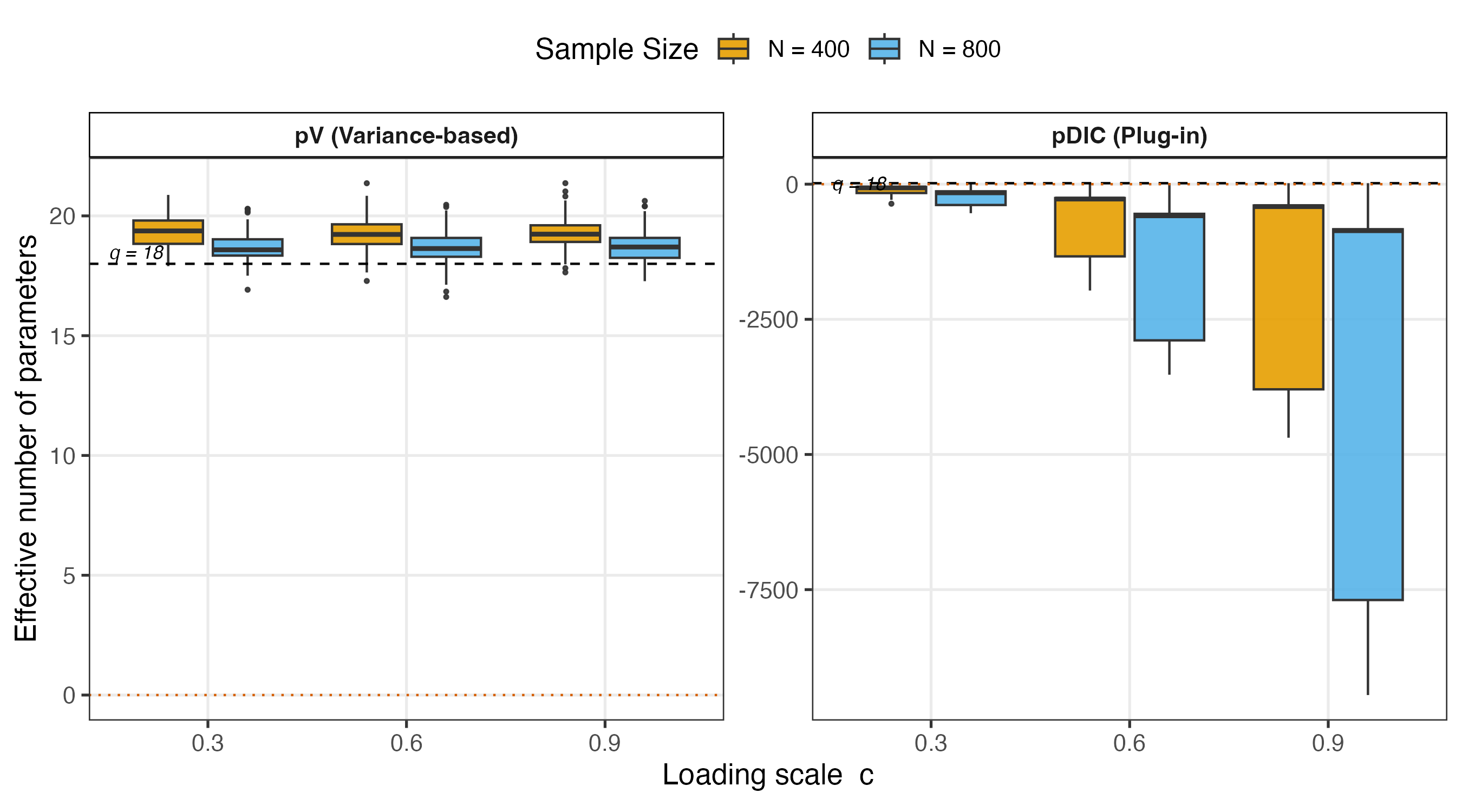}
    \caption{Box plots of variance-based and plug-in estimates of effective number of parameters by loading scale $c$ and sample size $J$, each comprising 200 replicates (100 each for $\sigma^2=1$ and $\sigma^2=0.5$).}
    \label{fig:FA_penalty_stability}
\end{figure}

\paragraph{Effective number of parameters.}
Table~\ref{tab:FA_penalty} presents summary statistics for the different estimates of the effective number of parameters across simulation conditions and replications. As expected due to sign switching,
 the classic plug-in estimate $\pDIC$ takes on extreme negative values and exhibits severe instability, with a mean of $-1{,}307.1$, standard deviation of $2{,}121.0$, minimum of $-9{,}450.1$ and maximum of  $18.2$.
The box-plots in the right panel of Figure~\ref{fig:FA_penalty_stability}
also show that $\pDIC$ exhibits extreme variability and
negativity, and that this behavior becomes more extreme as the true factor loadings increase (from $c=0.3$ to $c=0.9$). The likely reason for this behavior is that posterior mean loadings, pulled towards zero due to sign switching, are further from the true loadings and hence yield larger plug-in deviances (and therefore more negative $\pDIC$) when the true loadings are large than when they are closer to zero. Furthermore, differences in sign-switching behavior, i.e. two chains in each sign mode or one chain in one mode and three in the other, will have a larger impact on the plug-in deviance for larger $c$  and therefore lead to greater variability in $\pDIC$ across replicates.

Figure~\ref{fig:FA_penalty_stability} also shows that the variance-based estimate $p_V$ is much more stable and near the parameter count
$q = 18$ (dashed line). (The box-plots are marginal over the unique factor variance -- the full six-panel version separating $\sigma^2$
levels is provided in Figure~\ref{fig:supp_penalty_full} of the Supplementary Material.)
% ---------------------------------------------------------------
% TABLE: FA PENALTY STATISTICS
% ---------------------------------------------------------------
\begin{table}[htbp]
\centering
\small
\caption{Summary statistics of effective number of parameters across 12 simulation conditions with 100 replicates per condition.}
\label{tab:FA_penalty}
\begin{tabular}{lrrrrr}
\toprule
\textbf{Penalty Term} & \textbf{Mean} & \textbf{SD} & \textbf{Median} & \textbf{Min} & \textbf{Max} \\
\midrule
$\pDIC$ (Plug-in)      & $-1{,}307.1$ & $2{,}121.0$ & $-399.3$ & $-9{,}450.1$ & $18.2$ \\
$\pV$ (Variance-based)  & $19.0$       & $0.7$       & $18.9$   & $16.6$       & $21.4$ \\
$\pWAIC$      & $17.4$       & $0.4$       & $17.3$   & $16.2$       & $18.4$ \\
$\pLOO$       & $17.4$       & $0.4$       & $17.4$   & $16.2$       & $18.5$ \\
\bottomrule
\end{tabular}
\end{table}
% ---------------------------------------------------------------
% FIGURE: FA PENALTY STABILITY
% ---------------------------------------------------------------

Returning to Table~\ref{tab:FA_penalty}, the variance-based estimate $\pV$ and the \WAIC\ and LOO-CV estimates $\pWAIC$ and $\pLOO$ of the effective number of parameters are much more stable than $\pDIC$ and have means ($19.0$, $17.4$, and $17.4$, respectively) close to the number of free parameters ($q = 18$), consistent with the theoretical result that $\pV$ and $\pWAIC$  converge to $q$ in regular models.

When sign-switching did not occur ($\pDIC > 0$; 11\% of all 1{,}200 replications), both DIC variants agreed closely with \WAIC, with a root mean squared difference (RMSD) between DIC variant and WAIC of  $3.21$ for $\DICpVtwo$, and
$1.73$ for $\DICpV$. This confirms that the failure of the classic DIC is due to sign switching and that a negative $\pDIC$ is a useful diagnostic for such switching behavior.

\paragraph{Alignment of DIC variants with WAIC.}
Table~\ref{tab:FA_results_summary} presents RMSDs between each of the DIC variants  and the \WAIC\ for each of the 12 simulation
conditions. To establish a baseline, we first computed the root
mean squared difference (RMSD) between \WAIC and LOO-CV.
(column~5). As expected from their asymptotic equivalence,
\WAIC and \LOO are nearly identical, with an RMSD of $0.04$
across all conditions.

\begin{table}[htbp]
\centering
\small
\caption{Comparison of information criteria for the 12 simulation
conditions (rows, 100 replicates each).
RMSD (relative to WAIC) is the root mean squared difference between each information criterion and the  \WAIC. The last column is the  RMSD for  $\DICpV$ (column 6) divided by the standard deviation of the \WAIC (column 4).}
\label{tab:FA_results_summary}
\begin{tabular}{cccrrrrrr}
\toprule
\multirow{2}{*}{$c$} & \multirow{2}{*}{$\sigma^2$} & \multirow{2}{*}{$J$} & \multirow{2}{*}{SD of \WAIC} & \multicolumn{4}{c}{RMSD (relative to \WAIC) for:} & \multirow{2}{*}{\shortstack{RMSD for $\DICpV$ \\ $\div$ SD of \WAIC}} \\
\cmidrule(lr){5-8}
& & & & LOO-CV & $\DICpV$ & $\DICpVtwo$ & DIC & \\
\midrule
0.3 & 0.5 & 400  &  79.6 & 0.039 & \textbf{2.15} &  142 &  139 & \textbf{0.027} \\
0.3 & 0.5 & 800  &  91.9 & 0.037 & \textbf{1.14} &  290 &  288 & \textbf{0.012} \\
0.3 & 1.0 & 400  &  70.7 & 0.040 & \textbf{2.20} &  141 &  137 & \textbf{0.031} \\
0.3 & 1.0 & 800  & 106.8 & 0.037 & \textbf{1.22} &  299 &  297 & \textbf{0.011} \\
\addlinespace
0.6 & 0.5 & 400  &  69.7 & 0.041 & \textbf{2.07} &  938 &  935 & \textbf{0.030} \\
0.6 & 0.5 & 800  &  99.4 & 0.037 & \textbf{1.26} & 1{,}977 & 1{,}976 & \textbf{0.013} \\
0.6 & 1.0 & 400  &  68.1 & 0.042 & \textbf{2.01} &  882 &  880 & \textbf{0.030} \\
0.6 & 1.0 & 800  & 104.1 & 0.037 & \textbf{1.17} & 1{,}908 & 1{,}906 & \textbf{0.011} \\
\addlinespace
0.9 & 0.5 & 400  &  66.5 & 0.042 & \textbf{1.93} & 2{,}540 & 2{,}537 & \textbf{0.029} \\
0.9 & 0.5 & 800  &  99.3 & 0.038 & \textbf{1.23} & 5{,}239 & 5{,}237 & \textbf{0.012} \\
0.9 & 1.0 & 400  &  72.6 & 0.042 & \textbf{2.05} & 2{,}436 & 2{,}433 & \textbf{0.028} \\
0.9 & 1.0 & 800  &  89.8 & 0.038 & \textbf{1.15} & 5{,}082 & 5{,}081 & \textbf{0.013} \\
\bottomrule
\end{tabular}
\end{table}

The proposed \DICpV does not track the \WAIC quite as closely but the
RMSD is small, ranging from $1.14$ to $2.20$, with a
clear dependence on sample size: the RMSD roughly halves when
$J$ doubles (e.g., from ${\sim}2.1$ at $J = 400$ to
${\sim}1.2$ at $J = 800$), consistent with asymptotic equivalence established in
Section~\ref{subsec:dicpv_waic_relation}. The
loading scale $c$ and residual variance $\sigma^2$ have
little effect on the  discrepancy between $\DICpV$ and $\WAIC$.

In contrast to $\DICpV$, both $\DICpVtwo$ and DIC
have extremely large RMSDs, ranging from 137 to 5,082, and the RMSDs increase dramatically with
both loading scale and sample size. At $c = 0.9$ and
$J = 800$, both exhibit RMSD values exceeding $5{,}000$, i.e., an
increase from the $J = 400$ values of ${\sim}2{,}500$. This
worsening occurs because the deviance at a given distance from the mode (e.g. when the posterior mean loadings are close to 0, whereas the mode is near the true loadings) increases with sample size.
As shown in Section 4.1, $\DIC$ and $\DICpVtwo$ differ from $\DICpV$, in opposite directions, by $\pV-\pDIC$, which is more than about 19 here when $\pDIC<0$. Since  $\DICpV$ tends to be close to $\WAIC$, with an RMSD below 2.2, we also find that $\DIC$ and $\DICpVtwo$ deviate from \WAIC in opposite directions when sign switching occurs, as confirmed
by a near-perfect negative correlation for replicates with negative $\pDIC$.

To contextualize the magnitude of the discrepancy between  \DICpV and \WAIC, we divided
the corresponding RMSD by the within-condition standard deviation  of the \WAIC.
The resulting ratios, shown in the last column of Table~\ref{tab:FA_results_summary}, range from $0.011$ (at $J = 800$) to $0.031$ (at $J = 400$), indicating that  \DICpV as a reliable proxy for the \WAIC that improves with sample size.

\begin{figure}[htbp]
  \centering
  \includegraphics[width=.7\linewidth]{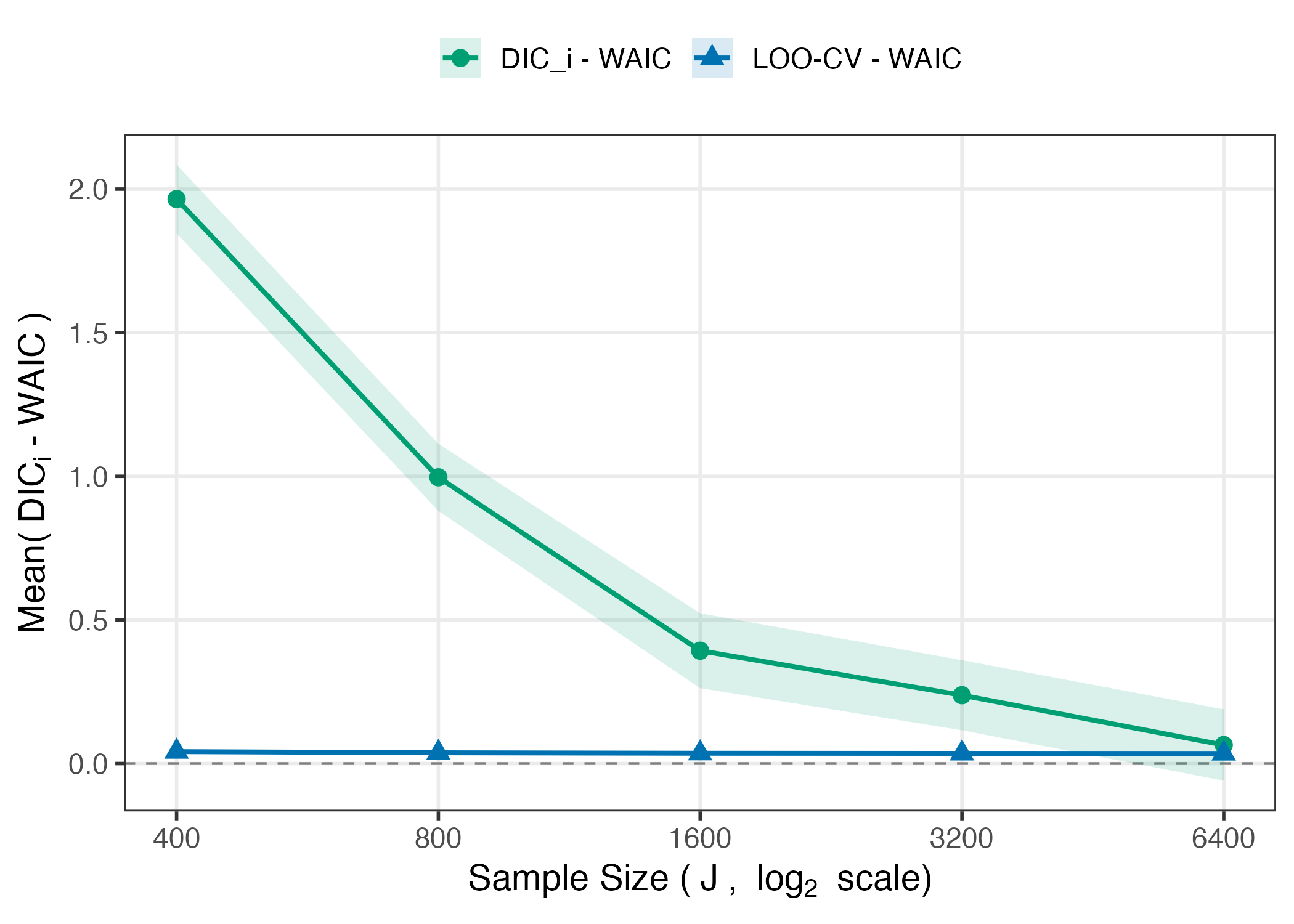}
  \caption{Mean difference $\DICpV - \WAIC$ as a function
of sample size ($J = 400$ to $6{,}400$, $\log_2$ scale)
for $c = 0.9$, $\sigma^2 = 1.0$, with 100 replicates
per sample size. Shaded ribbons denote 95\% confidence
intervals.}
  \label{fig:FA_convergence}
\end{figure}

\paragraph{Asymptotic convergence.}
To verify empirically that  \DICpV and \WAIC converge to the same limit, we extended the
simulation for one condition ($c = 0.9$, $\sigma^2 = 1.0$) to
sample sizes $J = 1{,}600$, $3{,}200$, and $6{,}400$ with 100
replicates each. Figure~\ref{fig:FA_convergence} displays the
mean difference $\DICpV - \WAIC$ as a function of $J$ on a $\log_2$ scale, with 95\% confidence intervals. The mean difference
decreases monotonically from approximately $2.0$ at $J = 400$
to $0.06$ at $J = 6{,}400$, with the confidence interval
covering zero at the largest sample size, consistent with the asymptotic result. Correspondingly, the variance-based penalty $\pV$ converges toward the parameter count $q = 18$ as $J$ increases. At $J = 6{,}400$, the mean of $\pV$ across the 100 replicates was 18.05, and the standard deviation was 0.66.

\paragraph{Computational Efficiency.}
As summarized in Table~\ref{tab:FA_timing}, \DICpV\ offers a substantial computational advantage. The median evaluation time for \DICpV\ was 1.55~ms per dataset, compared to 26.9~ms for \WAIC\ ($17\times$ slower) and 217.0~ms for LOO-CV ($140\times$ slower). This efficiency stems from \DICpV's use of the pre-computed joint deviance, whereas the Bayesian criteria require operations on the $J \times S$ matrix of pointwise likelihoods.
\begin{table}[htbp]
\centering
\caption{Median computation times and relative speedups for information criteria across all 1{,}200 fitted models.}
\label{tab:FA_timing}
\begin{tabular}{lccc}
\toprule
Criterion & Median time (ms) & \DICpV vs WAIC & \DICpV vs LOO \\
\midrule
\DICpV       & 1.55  & ---          & --- \\
WAIC         & 26.9  & 17.7$\times$ & --- \\
LOO-CV    & 217.0 & ---          & 139.5$\times$ \\
\bottomrule
\end{tabular}
\end{table}

%--------------------------------------------------------------------------
% Growth Mixture Models
%--------------------------------------------------------------------------
\subsection{Simulation Study II: GMMs with Parameterization Switching}
\label{sec:gmm}

\subsubsection{Simulation Design}
\label{sec:gmm_simdesign}
\begin{figure}[htbp]
  \centering
  \includegraphics[width=\linewidth]{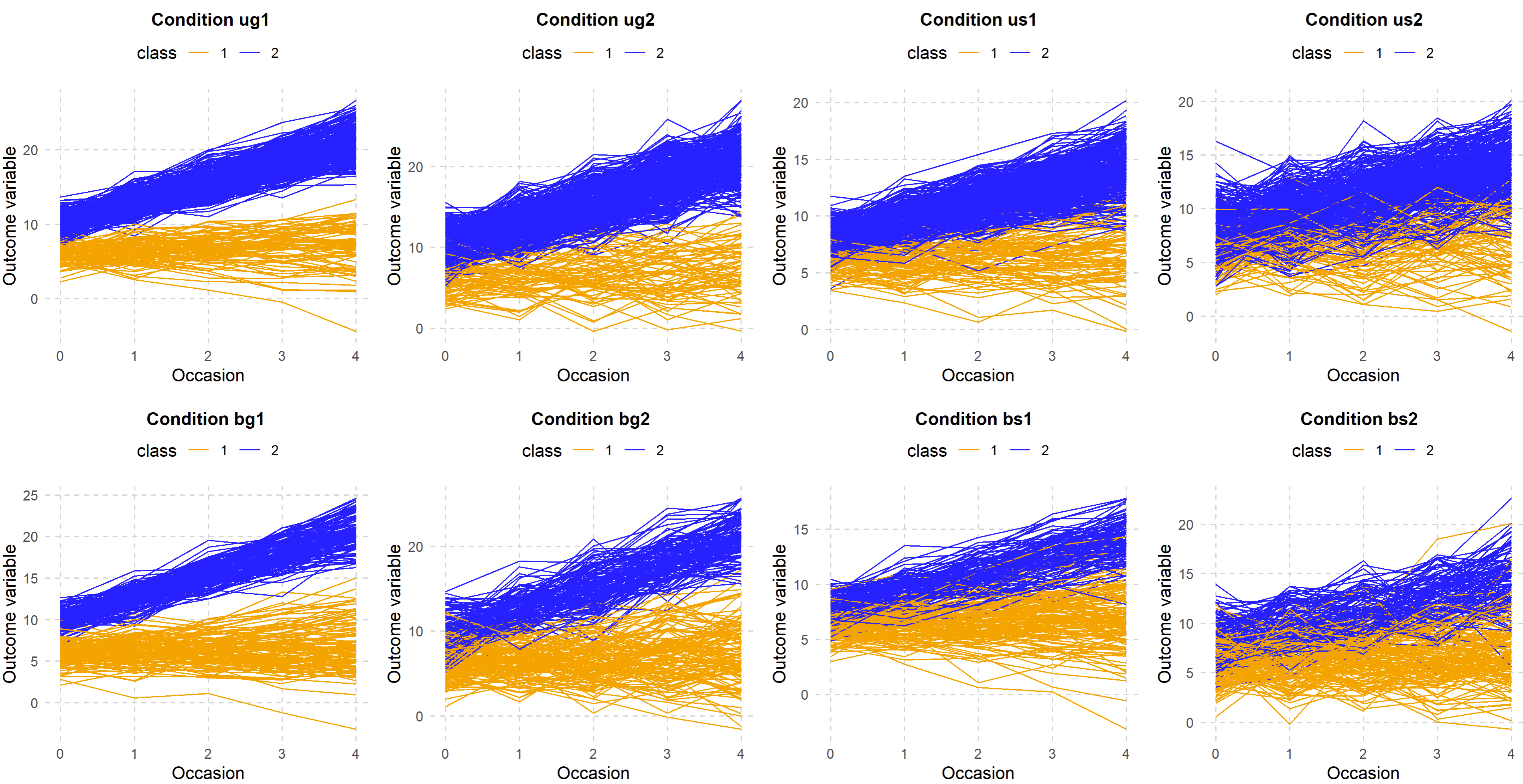}
  \caption{Subject trajectories by class across the eight simulation conditions. \textit{Note}. Rows distinguish residual noise levels; columns vary by class balance and separation.}
  \label{fig:sim_trajectories}
\end{figure}

We simulated data from the model defined in Section~\ref{subsec:model_gmm}, with $K=2$  classes  and five equidistant time points ($n_j=5$, $t_{1j}=0,\ldots, t_{5j}=4$). We set the quadratic term to zero, $\beta_2^{(k)}=0$, for both classes, reducing the trajectories to linear growth. The class-specific covariance matrices $\bvec{\Psi}^{(k)}$ for the varying intercept and slope of $t_{ij}$ were held constant across conditions. Letting $\sigma_1^{(k)}$ and $\sigma_2^{(k)}$ denote the intercept and slope standard deviations and $\rho_{12}^{(k)}$ their correlation, we set $\sigma_1^{(1)}=0.8$, $\sigma_2^{(1)}=0.6$, $\rho_{12}^{(1)}=0.2$ for class~1 and $\sigma_1^{(2)}=0.5$, $\sigma_2^{(2)}=0.3$, $\rho_{12}^{(2)}=0.8$ for class~2, with class~1 having larger variances and weaker correlation. We utilized a $2 \times 2 \times 2$ factorial design:
\begin{enumerate}[label=(\roman*), nosep]
    \item \textbf{Class balance:} Balanced ($\pi^{(1)}=0.5$, denoted \texttt{b}) with $J=250$ vs.\ Unbalanced ($\pi^{(1)}=0.2$, denoted \texttt{u}) with $J=400$;
    \item \textbf{Class separation:} Greater (denoted \texttt{g}) vs.\ smaller (denoted \texttt{s}) separation in growth trajectories.  The class-specific mean (or fixed) intercepts and slopes are $(\beta_0^{(1)}, \beta_0^{(2)}) = (6, 10)$ and $(\beta_1^{(1)}, \beta_1^{(2)}) = (0.3, 2.7)$ for greater separation, and $(6, 8)$ and $(0.3, 1.5)$ for smaller separation; see \citet[\S 3.3.2]{xiao:2025} for additional details on the parameter choices;
    \item \textbf{Residual noise:} Low ($\sigma_e=1$) vs.\ High ($\sigma_e=2$).
\end{enumerate}
This resulted in eight conditions, named by combining the characters \texttt{b} versus \texttt{u} for class balance with \texttt{g} versus \texttt{s} for class separation, and \texttt{1} versus \texttt{2} for residual noise (e.g., \texttt{us2} denotes unbalanced, small separation, high noise). We generated 50 replicate datasets per condition. Candidate models with $K \in \{1, 2, 3, 4\}$ classes were fitted for each dataset, yielding 1{,}600 fitted models in total. The number of free parameters is $q=7K$ per model: $(K-1)$ mixing weights, $3K$ class-specific fixed regression coefficients, $3K$ class-specific covariance parameters, and 1 residual variance.  Figure~\ref{fig:sim_trajectories} shows simulated trajectories for each condition.

\subsubsection{Estimation and Computation}
Candidate models were fitted using \texttt{CmdStan} \citep{Stan2021} (4 chains, 1{,}000 warmup and 1{,}000 post-warmup draws each). Following the prior specification in
Section~\ref{subsec:model_gmm}, we set the Dirichlet
concentration parameter to $\alpha = 10$, standard deviation priors to
$\text{Half-Normal}(0, 10)$, and correlation prior to
$\text{LKJ}(2)$. As discussed in Section~\ref{subsec:identifiability} where we cite \citet{rousseau:2011}, with $\alpha = 10$ we expect that classes will tend to merge in overfitted models rather than vanish.

We corrected for label switching post-hoc using Kullback--Leibler divergence minimization \citep{stephens:2000}. Although this was not necessary for the parameterization-invariant information criteria
(\WAIC\, LOO-CV, and \DICpV), we corrected for label switching so that we could attribute poor performance of DIC and \DICpVtwo to  other issues, such as parameterization switching.

\subsubsection{Results}
\label{sec:gmm_results}

\paragraph{Effective number of parameters.}
Table~\ref{tab:GMM_penalty} summarizes the distribution of the estimated effective number of parameters across all 1{,}600 fitted models. The effective number of parameters  $\pDIC$ for the classic DIC had a much larger standard deviation than the others (SD $= 20.3$ versus 3.6 to 5.5) and became negative (minimum $= -166.4$), as expected if parameterization switching occurs.
% ---------------------------------------------------------------
% TABLE: GMM PENALTY STATISTICS
% ---------------------------------------------------------------
\begin{table}[htbp]
\centering
\small
\caption{Distribution of effective number of parameters across 1{,}600 GMM fitted models ($K = 1, \dots, 4$, 8 conditions, 50 replicates).}
\label{tab:GMM_penalty}
\begin{tabular}{lrrrrr}
\toprule
\textbf{Effective number of parameters} & \textbf{Mean} & \textbf{SD} & \textbf{Median} & \textbf{Min} & \textbf{Max} \\
\midrule
$\pDIC$ (Plug-in)         & $10.4$ & $20.3$ & $16.6$ & $-166.4$ & $23.8$ \\
$\pV$ (Variance-based)  & $13.7$  & $5.5$  & $13.8$  & $5.5$    & $49.1$ \\
$\pWAIC$      & $10.7$ & $3.6$  & $11.0$ & $4.0$    & $20.5$ \\
$\pLOO$       & $10.8$ & $3.6$  & $11.0$ & $4.0$    & $20.6$ \\
\bottomrule
\end{tabular}
\end{table}
In contrast, the effective number of parameters for the other criteria  remained strictly positive, with minimum values of 5.5 for $\pV$ and 4.0 for  \WAIC\ and LOO-CV.

Negative $\pDIC$ occurred only in the greater separation conditions (\texttt{ug1}, \texttt{ug2}, \texttt{bg1}, \texttt{bg2}) and only when $K=3$ or $K=4$, i.e., when the specified number of classes was greater than the true
number of classes. As shown in Table 5,  the largest classic
Gelman--Rubin $\widehat{R}$ \citep{Gelman:92} across parameters, denoted $\Rhatm$, almost always exceeded 1.1 when $\pDIC$ was negative. For example, for the 50 datasets generated under condition \texttt{ug1}, fitting
4-class GMMs resulted in negative $\pDIC$ for 32 datasets and for 31 of these, $\Rhatm >1.1$.
\begin{table}[htbp]
\centering
\small
\caption{Number of replicate datasets with negative $\pDIC$ (number of these datasets for which $\Rhatm> 1.1$) across 50 replicates per condition, separately for specified models with $K=3$ and $K=4$. }
\label{tab:GMM_pDIC_Rhat}
\begin{tabular}{lcc}
\toprule
Condition & $K=3$ & $K=4$ \\
\midrule
\texttt{ug1} & 26 (25) & 32 (31) \\
\texttt{ug2} &  \rule{1.5mm}{0mm}9 \rule{1.7mm}{0mm}(8)  & 11 (10) \\
\texttt{bg1} & 10 (10) & 30 (29) \\
\texttt{bg2} & 13 (13) & 14 (12) \\
\bottomrule
\end{tabular}
\end{table}

\begin{figure}[htbp]
  \centering
  \includegraphics[width=\linewidth]{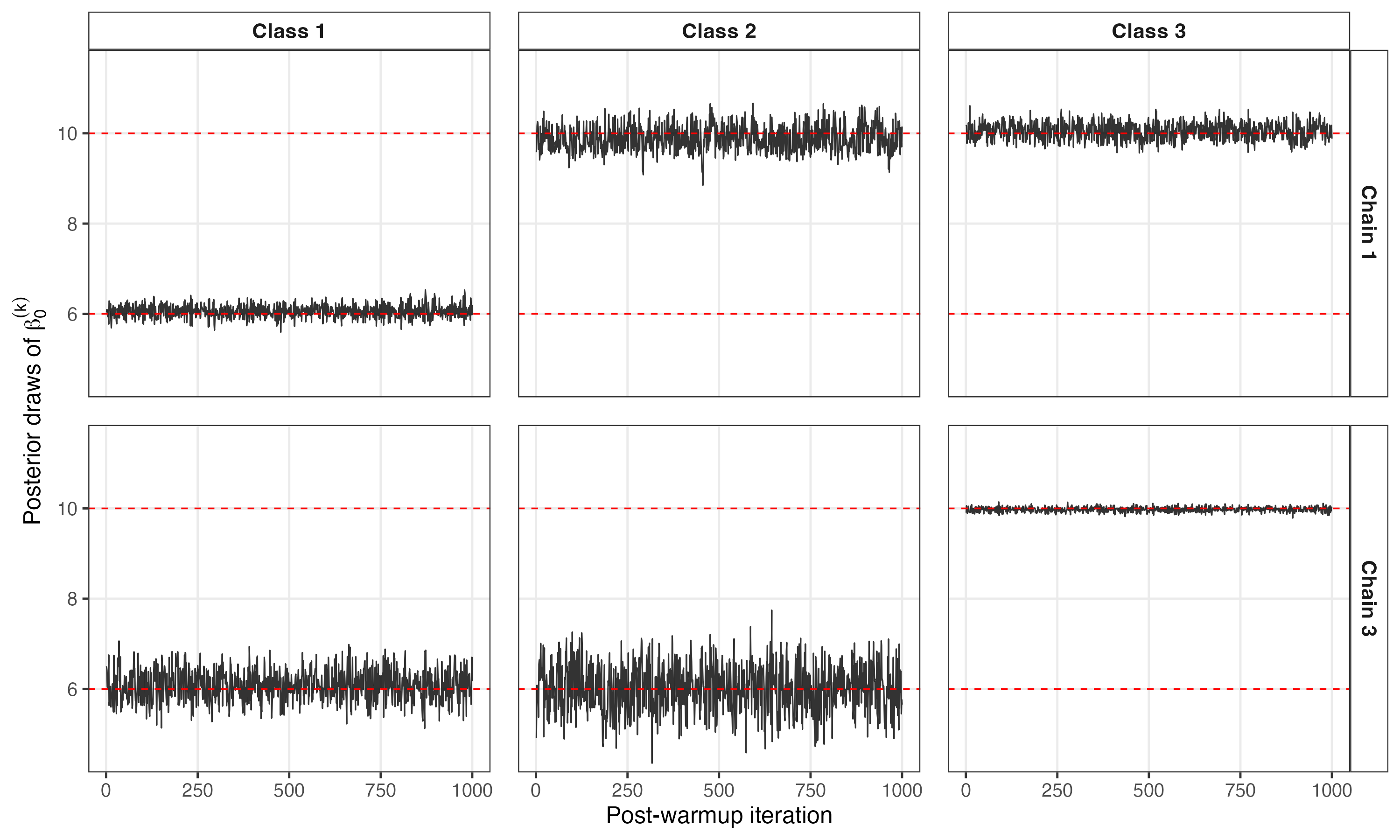}
  \caption{Traceplots for two chains (rows) of the class-specific intercepts $\beta_0^{(1)}$,  $\beta_0^{(2)}$, and  $\beta_0^{(3)}$  (columns) for a 3-class GMM fitted to replicate~3 of condition \texttt{ug1}.}
  \label{fig:GMM_traceplot}
\end{figure}

These findings can be explained by parameterization switching occurring largely between chains, as illustrated  in Figure~\ref{fig:GMM_traceplot} for a 3-class model fitted to replicate~3 in condition \texttt{ug1}.
For Chains 1 and 3 (rows), traceplots of the intercept parameter are shown for each of the three specified classes (columns). The two-class  data-generating model has intercepts equal to $6$ and $10$, shown as dashed lines. Whereas $\beta_0^{(2)}$ (middle column) settles near $10$ in Chain~1 (top row), it settles near $6$ in  Chain~3 (bottom row). This is because the specified three-class model (nearly) degenerates to a two-class model in two ways: the second class merges with the third class ($\beta_0^{(2)}$ near 10) in Chain 1 and it merges with the first class ($\beta_0^{(2)}$ near 6) in Chain 3.
\citet[Web-appendix~F]{Xiao_Rabe-Hesketh_Skrondal_2025} provide further evidence that the ``redundant'' class 2 merges with one of the classes in one chain and with the other class in the other chain in this replicate.
\citet[\S3.7.1]{xiao:2025} looked more closely at the MCMC chains
where $\pDIC<0$ and $\Rhatm<1.1$ and found that within-chain switching occurred.  For example, for a replicate in condition \texttt{bg1}, the model with $K=4$ frequently switched between two parameterizations within each chain, whereas
for a replicate in condition
 \texttt{ug2}, switching occurred for less than 200 iterations in one of four chains for a model with $K=3$.

\begin{figure}[htbp]
  \centering
  \includegraphics[width=\linewidth]{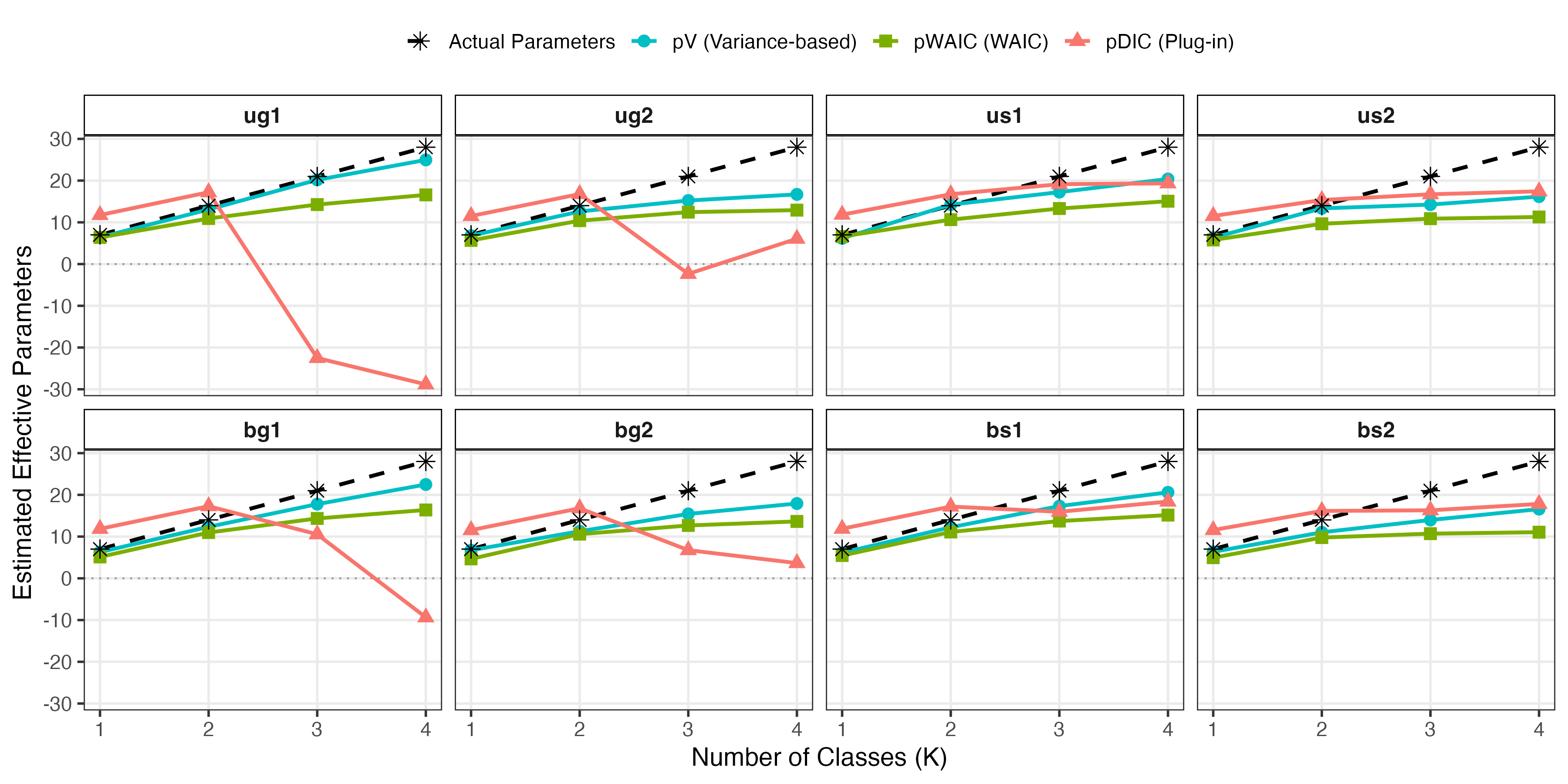}
  \caption{Parameter count $q=7K$ (black, dashed line, asterisks) and means (over 50 replicate datasets) of the effective number of parameters  $\pV$ (teal, circles), $\pWAIC$ (green, squares)  and $\pDIC$ (red, triangles) versus number of classes ($K$) by simulation condition.}
  \label{fig:GMM_penalty_comp}
\end{figure}

Figure~\ref{fig:GMM_penalty_comp} shows that the averages (over replicate datasets) of both $\pV$ and $\pWAIC$ increased monotonically with the number of classes~$K$ across all conditions,  but fell below the  parameter count $q=7K$ (dashed line), possibly due to relatively strong prior information.
In contrast, the average $\pDIC$ decreased when $K$ increased from 2 in the four large-separation conditions where parameterization-switching occurred, becoming negative for \texttt{ug1} and \texttt{bg1}.

% ---------------------------------------------------------------
% FIGURE: GMM PENALTY COMPARISON
% ---------------------------------------------------------------

\paragraph{Alignment of DIC variants with \WAIC.}
Table~\ref{tab:GMM_results_summary} compares the \DIC variants and \WAIC.
The proposed \DICpV is quite close to the \WAIC\ with RMSDs ranging from $2.8$ to $5.7$  across simulation conditions. While these RMSDs are larger than those for the LOO-CV, which do not exceed 0.15, they are  much smaller
% ---------------------------------------------------------------
% TABLE: GMM COMBINED RESULTS (RMSD & CONTEXT)
% ---------------------------------------------------------------
\begin{table}[htbp]
\centering
\small
\caption{Comparison of information criteria across 8 simulation conditions (rows; 50 replicates per condition $\times$ 4 candidate models). RMSD (relative to \WAIC) is the root mean squared difference between each information criterion and \WAIC. The last column is the RMSD for $\DICpV$ (column 4) divided by the within-model SD of \WAIC (column 2).}
\label{tab:GMM_results_summary}
\begin{tabular}{ccrrrrc}
\toprule
\multirow{2}{*}{Condition} & \multirow{2}{*}{SD of \WAIC} & \multicolumn{4}{c}{RMSD (relative to \WAIC) for:} & \multirow{2}{*}{\shortstack{RMSD for $\DICpV$ \\ $\div$ SD of \WAIC}} \\
\cmidrule(lr){3-6}
& & LOO-CV & $\DICpV$ & $\DICpVtwo$ & DIC & \\
\midrule
\texttt{ug1} & 79.2 & 0.13 & \textbf{5.7} & 53.3 & 45.8 & \textbf{0.071} \\
\texttt{ug2} & 60.8 & 0.05 & \textbf{2.9} & 29.7 & 26.7 & \textbf{0.048} \\
\texttt{us1} & 72.4 & 0.09 & \textbf{3.7} &  5.2 &  5.4 & \textbf{0.052} \\
\texttt{us2} & 66.0 & 0.05 & \textbf{3.8} &  4.6 &  5.9 & \textbf{0.058} \\
\texttt{bg1} & 51.0 & 0.15 & \textbf{3.8} & 26.6 & 22.1 & \textbf{0.074} \\
\texttt{bg2} & 45.5 & 0.06 & \textbf{2.8} & 17.9 & 15.3 & \textbf{0.061} \\
\texttt{bs1} & 50.3 & 0.12 & \textbf{3.3} &  6.5 &  5.0 & \textbf{0.066} \\
\texttt{bs2} & 52.0 & 0.15 & \textbf{3.4} &  4.2 &  6.3 & \textbf{0.065} \\
\bottomrule
\end{tabular}
\end{table}
than the RMSDs for the \DICpVtwo and \DIC in the conditions where switching occurs (\texttt{ug1}, \texttt{ug2}, \texttt{bg1}, \texttt{bg2}).
In these four conditions with greater class separation, \DICpVtwo\ and DIC have RMSDs ranging from 15.3 to 53.3. The RMSDs for the \DICpVtwo\ and DIC are similar across conditions. This is because they deviate from $\DICpV$ by the same amount $\pV-\pDIC$, in opposite directions, and
$\DICpV$ is fairly close to \WAIC.
To contextualize the RMSD for the \DICpV, we divided it by the within-model standard deviation of the \WAIC  (in the second column), giving the ratios in the last column of the table. Here the within-model standard deviation of the \WAIC was  obtained by computing the variance for each of the four candidate models ($K=1$ to $K=4$), averaging the variances, and taking the square root. This isolates the sampling variability for a fixed model from between-model variability. As shown in the last column of the table, these ratios range from $0.048$ to $0.074$, suggesting that the \DICpV is a reasonable proxy for the WAIC.

\begin{figure}[htbp]
  \centering
  \includegraphics[width=\linewidth]{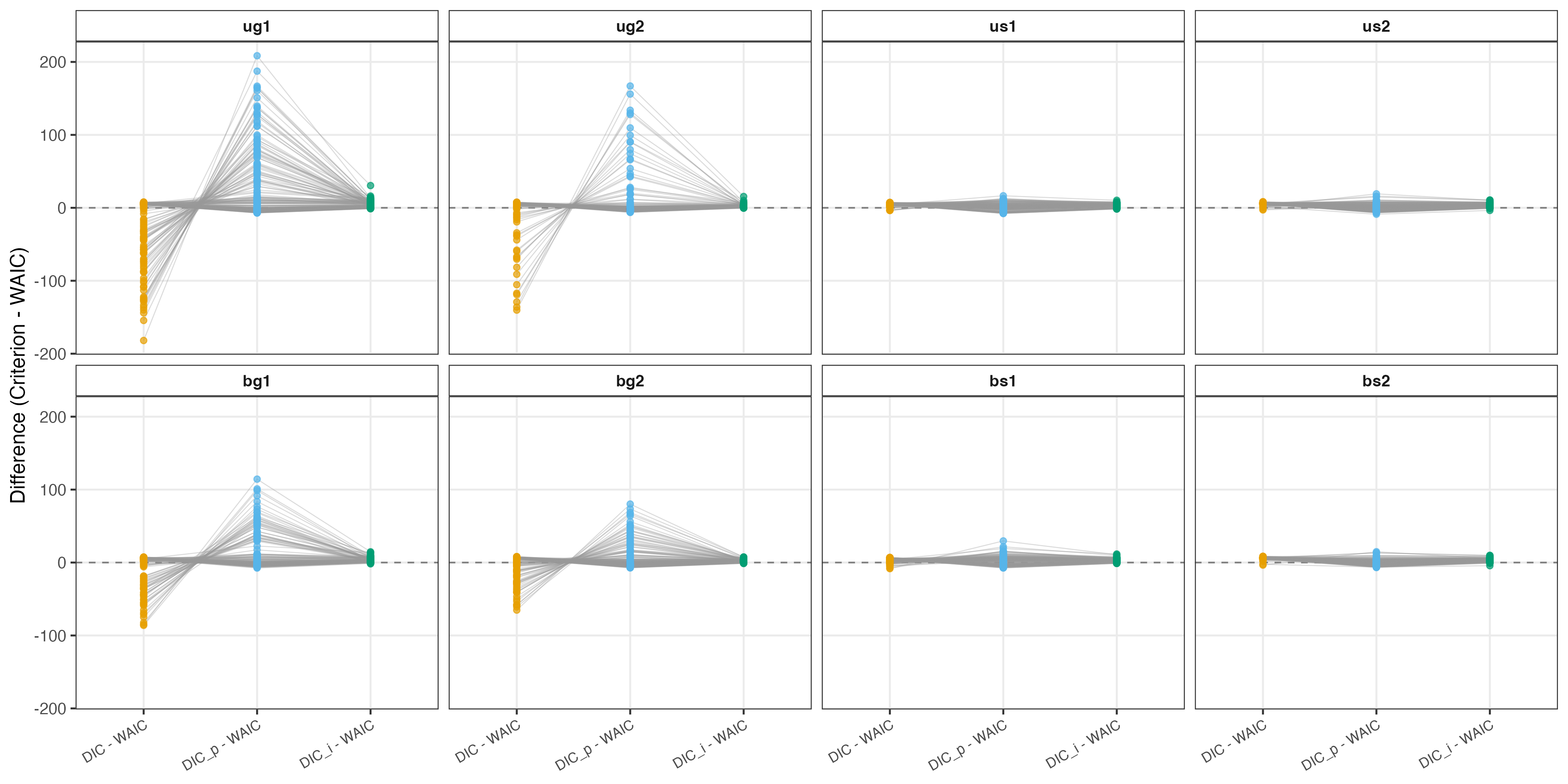}
  \caption{Difference (Criterion $-$ \WAIC) by replicate and condition. Within each panel, the criteria are arranged from left to right: classic DIC (yellow), \DICpVtwo (blue), and \DICpV (green). }
  \label{fig:GMM_alignment}
\end{figure}

Figure~\ref{fig:GMM_alignment} shows the differences between each DIC variant and the WAIC (y-axis), for DIC (left on x-axis, yellow), $\DICpVtwo$ (middle on x-axis, blue), and $\DICpV$ (right on x-axis, green) for each condition. Points for the same replicate and model are connected. Whereas  \DICpV$-$WAIC differences  cluster tightly around zero, \DICpVtwo shows a systematic positive shifts in the four greater separation conditions  that approximately mirror the negative shifts of  DIC.
% ---------------------------------------------------------------
% FIGURE: GMM ALIGNMENT WITH WAIC
% ---------------------------------------------------------------

\begin{figure}[htbp]
  \centering
  \includegraphics[width=\linewidth]{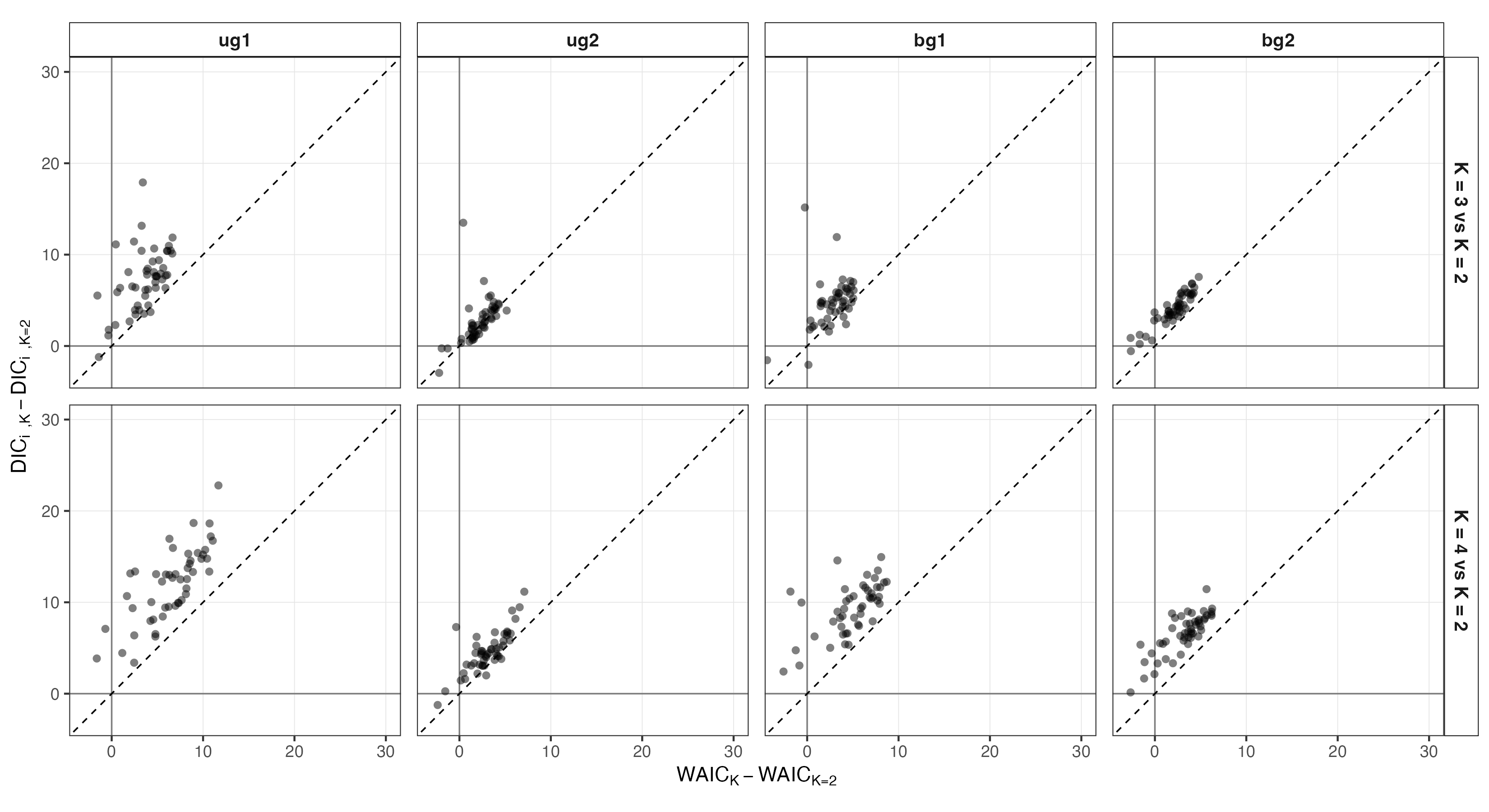}
  \caption{Differences in $\DICpV$ versus differences in \WAIC\ for overfit candidate models relative to the true model ($K=2$), under the four greater-separation conditions. Top row: $K=3$ versus $K=2$. Bottom row: $K=4$ versus $K=2$. Each point represents one of 50 replicate datasets per condition. The dashed line is $y = x$. Solid gray lines mark the origin. Points in the upper-right quadrant indicate that both criteria favor the true 2-class model.}
  \label{fig:GMM_model_comparison}
\end{figure}

\paragraph{Model comparison.}
To assess the performance of the different information criteria for model comparison, we computed the difference of each criterion between an overfit candidate model ($K=3$ or $K=4$) and the true model ($K=2$) for each replicate dataset; a positive difference means that the criterion correctly assigns a worse value to the overfit model. Figure~\ref{fig:GMM_model_comparison} shows scatterplots of $\DICpV$ differences against the corresponding \WAIC differences for the four greater separation models, revealing close tracking, with most points falling near the $y = x$ line and in the upper-right quadrant where both criteria favor the true model. Across these conditions, $\DICpV$ selects $K=2$ in 94--100\% of replicates and \WAIC in 84--96\%, depending on condition and whether 3-class or 4-class model is being compared with the true model. Figure~\ref{fig:supp_3criteria_greater} in the supplementary materials presents the same figure but with all three DIC variants overlaid, revealing that \DICpVtwo-differences often lie far above the $y = x$ line, whereas the DIC-differences tend to lie below the line by a similar amount. Figure~\ref{fig:supp_3criteria_smaller} in the supplementary material shows the corresponding plot for the four smaller-separation conditions, where all three criteria agree closely with \WAIC. Finally, Figure~\ref{fig:supp_3criteria_K1} in the supplementary material shows the underfit comparison ($K=1$ versus $K=2$), where all three criteria agree quite closely across all eight conditions, consistent with the poor performance of \DICpVtwo and DIC in overfit models being due to degenerate nonidentifiability.

%==========================================================================
% SECTION 6: DISCUSSION
%==========================================================================
\section{Discussion}
\label{sec:discussion}

We introduced \DICpV, a version of the DIC that does not rely on the plug-in deviance which often becomes unstable and so large that the classic DIC penalty is negative. Unlike the classic DIC, \DICpV is also invariant to reparameterization. This is useful when transformations of parameters are arbitrary, for example when choosing between a standard deviation, log standard deviation, or variance. Importantly, invariance to reparameterization also means that \DICpV performs well when there are identifiability issues, examples being reflection invariance in factor analysis and degenerate nonidentifiability in finite mixture models. Whereas the classic DIC is known to perform poorly for finite mixture models and is therefore not provided for such models by some software (e.g., \texttt{OpenBUGS}),  we showed that \DICpV performs well for growth mixture models. Since  \DICpV is invariant to permutations of the class labels, its performance is not affected by label switching.

\citet{gelman2014understanding} proposed replacing the penalty term of the classic DIC by the posterior variance of the deviance, giving an alternative DIC denoted \DICpVtwo here. Our proposed \DICpV can be viewed as the average of the classic DIC and \DICpVtwo. When the plug-in deviance becomes unstable and extremely large, for example due to multimodality of the posterior, both DIC and \DICpVtwo become extreme, in opposite directions, whereas \DICpV remains stable.

We showed that \DICpV is asymptotically equivalent to the WAIC and found that the root mean squared difference between \DICpV and WAIC was small compared with the sampling standard deviation of the WAIC in our simulations.
When comparing growth mixture models with different numbers of classes, \DICpV\ and WAIC tended to choose the same model, with \DICpV\ slightly outperforming \WAIC\ at rejecting overfit models.

 \WAIC and LOO-CV can be viewed as the  criteria of choice for evaluating predictive accuracy of Bayesian models \citep[e.g.,][]{gelman2014understanding,vehtari:2016} because they are based on fully Bayesian posterior predictive densities. However, both  criteria rely on factorization of the likelihood into ``point'' contributions. When the likelihood is defined marginally over latent variables, the ``points'' are typically clusters of units, such as students belonging to the same schools in a multilevel model with varying school intercepts. However, some latent variable models do not permit any factorization of the marginal likelihood, for instance longitudinal models with latent variables for subjects and occasions. For such models, \WAIC and LOO-CV are not defined and \DICpV may be a good alternative. Computing \WAIC and LOO-CV is also computationally
 demanding because they require likelihood contributions for all ``points'' (units or clusters) at all parameter draws, whereas \DICpV requires only the overall likelihood or deviance evaluated at all parameter draws.

An R package \texttt{dicv} implementing the function presented in the appendix is available at \url{https://github.com/DoriaXiao/dicv}, with vignettes for Stan workflows. An interactive demonstration is available at \url{https://doriaxiao.shinyapps.io/dicv_app/}.

\citet{vehtari:2016} point out that the variance-based effective number of parameters used in \DICpV and \DICpVtwo  can be unstable, but we have not found that to be the case in the scenarios considered in our simulations. Unfortunately, there does not seem to be an obvious way to check the reliability of the \DICpV.
A possible approach for diagnosing problems could be to analyze the shape of the posterior deviance distribution and to compare chain-specific estimates of the posterior variances of the deviance across chains. For WAIC, \citet{vehtari:2016} recommend checking whether any of the posterior variances of the log point predictive densities exceed 0.4, and for LOO-CV they recommend checking whether the estimate of the shape parameter of the Pareto distribution exceeds 0.7. These diagnostics are also useful for identifying influential observations.

In summary, we recommend using \DICpV instead of the classic DIC, especially when the classic DIC breaks down by having a negative penalty term.
We also recommend computing the classic DIC penalty because a negative value is a signal that posterior means are poor estimates of the model parameters and that the classic DIC cannot be used. \DICpV will be particularly valuable for software that does not compute the WAIC or LOO-CV because, like DIC, \DICpV only requires deviances evaluated at the posterior parameter draws. For example, the Bayesian estimator in \texttt{Mplus} \citep{muthen2010bayesian} provides DIC, but not WAIC or LOO-CV, for finite mixture models including GMMs. It would take negligible programming to produce \DICpV as well, with the recommendation to use it either generally for finite mixture models or whenever $\pDIC$ is negative for any of the candidate models.  Finally, there are models for which  WAIC and LOO-CV are not available because the deviance does not factorize, and \DICpV may well be a good alternative for such models.

\small
\bibliographystyle{plainnat-full}
\bibliography{bibliography}

@article{muthen:2000,
  title={Integrating person-centered and variable-centered analyses: {G}rowth mixture modeling with latent trajectory classes},
  author={Muthén, B. O. and Muthén, L. K.},
  journal={Alcohol. Clinical \& Experimental Research},
  year={2000},
  volume={24},
  pages={882-891}
}

@MISC{CmdStan,
author = {{Stan Development Team}},
title = {Cmd{S}tan {U}ser’s {G}uide: Version 2.30},
year = {2021},
howpublished={\url{https://mc-stan.org/docs/cmdstan-guide/index.html}}
}

@article{diebolt:94,
    author= {Diebolt, Jean and Robert, Christian P.},
    year={1994},
    title={Estimation of finite mixture distributions through {B}ayesian
    sampling},
    journal={Journal of the Royal Statistical Society, Series B},
    volume={56},
    pages = {363-375}
}

@article{Gelman:92,
    author = {Andrew Gelman and Donald B. Rubin},
    title = {Inference from iterative simulation in multiple sequences (with discussion)},
    volume = {7},
    journal = {Statistical Science},
    pages = {457-511},
    year = {1992}
}

@book{gelman:2014,
  title={Bayesian Data Analysis},
  author={Andrew Gelman and John B. Carlin and Hal S. Stern and David B. Dunson and Aki Vehtari and Donald B. Rubin},
  edition={Third},
  publisher={Chapman \& Hall/CRC}, 
  address={Boca Raton},
  year={2014}
}

@article{gelman:96,
    author = {Andrew Gelman and Xiao-Li Meng and Hal Stern},
    title = {Posterior predictive assessment of model fitness via realized discrepancies},
    journal = {Statistica Sinica},
    volume = {6},
    pages = {733-807},
    year  = {1996}
}

@article{Merkle2019,
    author = {Merkle, E.C. and Furr, D. and Rabe-Hesketh, S.},
    year = {2019},
    pages = {802-829},
    title = {Bayesian Comparison of Latent Variable Models: Conditional Versus Marginal Likelihoods},
    volume = {84},
    journal = {Psychometrika}
}

@article{Merkle:2021,
    author = {Merkle, E.C. and Fitzsimmons, E. and Uanhoro, J. and Goodrich, B.},
    year = {2021},
    pages = {1-22},
    title = {Efficient {Bayesian} structural equation modeling in {Stan}},
    volume = {100},
    journal = {Journal of Statistical Software}
}

@article{Lewandowski:2009,
  title={Generating random correlation matrices based on vines and extended onion method},
  author={Lewandowski, Daniel and Kurowicka, Dorota and Joe, Harry},
  journal={Journal of Multivariate Analysis},
  year={2009},
  volume={9},
  pages={1989-2001}
}

@article{muthen:99,
  title={Finite mixture modeling with mixture outcomes using the {EM} algorithm},
  author={Muthén, B. O. and Shedden, K.},
  journal={Biometrics},
  year={1999},
  volume={55},
  pages={463-469}
}

@article{muthen:2002,
  title={Beyond {SEM}: {G}eneral latent variable modeling},
  author={Muthén, B. O.},
  journal={Behaviormetrika},
  year={2002},
  volume={29},
  pages={81-117}
}

@article{Pianta:2008,
    author = {Robert C. Pianta and Jay Belsky and Renate Houts and Fred J. Morrison},
    year = {2008},
    pages = {365-397},
    title = {Classroom Effects on Children’s Achievement Trajectories in Elementary School},
    volume = {45},
    journal = {American Educational Research Journal}
}

@article{plummer:2008,
    author={Martyn Plummer},
    year={2008},
    pages = {523-539},
    title = {Penalized loss functions for {Bayesian} model comparison},
    volume={9},
    journal = {Biostatistics}
}

@article{redner:84,
    author={Redner, R. A. and Walker, H. C.},
    year={1984},
    title={Mixture densities, maximum likelihood and the {EM} algorithm},
    journal={SIAM Review},
    volume={26},
    pages={195-239}
}

@misc{vehtari:2016,
    title={loo: Efficient leave-one-out cross-validation and {WAIC} for {Bayesian} models},
    note={{R} package version 0.1.6},
    author={Aki Vehtari and Andrew Gelman and Jonah Gabry},
    url={https://github.com/stan-dev/loo},
     year={2016}
}

@ARTICLE{vehtari2017,
  title={Practical {Bayesian} model evaluation using leave-one-out cross-validation and {WAIC}},
  author={Aki Vehtari and Andrew Gelman and Jonah Gabry},
  journal={Statistics and Computing},
  year={2017},
  volume={27},
  pages={1413-1432}
}

@Article{vehtari:2021,
    title = {Rank-normalization, folding, and localization: An improved
      $\widehat{R}$ for assessing convergence of {MCMC} (with discussion)},
    author = {Aki Vehtari and Andrew Gelman and Daniel Simpson and Bob
      Carpenter and Paul-Christian Bürkner},
    journal = {Bayesian Analysis},
    volume={16},
    number={2},
    pages={667--718},
    year={2021},
    publisher={International Society for Bayesian Analysis}
  }

@article{Watanabe2010,
  author = {Sumio Watanabe},
  title = {Asymptotic equivalence of {Bayes} cross validation and widely applicable information criterion in singular learning theory},
  journal = {Journal of Machine Learning Research},
  year = {2010},
  volume = {11},
  pages = {3571--3594}
}

@phdthesis{xiao:2025,
    author = {Xiao, Xingyao},
    title = {Bayesian Identification, Estimation, and Evaluation of Growth Mixture Models},
    school = {University of California, Berkeley},
    year = {2025}
}

@article{stacking, 
  author = {Yao, Yuling and Vehtari, Aki and Gelman, Andrew},
  title = {Stacking for Non-mixing {B}ayesian Computations: The Curse and Blessing of Multimodal Posteriors},
  pages = {1-45},
  volume = {23},
  journal = {Journal of Machine Learning Research},
  year = {2022}
}

@article{Spiegelhalter2002,
  title={Bayesian measures of model complexity and fit},
  author={David J. Spiegelhalter and Nicola G. Best and Bradley P. Carlin and Angelika van der Linde},
  journal={Journal of the Royal Statistical Society, Series B},
  year={2002},
  volume={64},
  pages={583-639}
}

@article{Kreuter2008AnalyzingCT,
  title={Analyzing Criminal Trajectory Profiles: Bridging Multilevel and Group-based Approaches Using Growth Mixture Modeling},
  author={Frauke Kreuter and Bengt Muth{\'e}n},
  journal={Journal of Quantitative Criminology},
  year={2008},
  volume={24},
  pages={1-31}
}

@MISC{Stan2021,
author = {{Stan Development Team}},
title = {Stan Reference Manual: Version 2.30},
year = {2021},
howpublished={\url{https://mc-stan.org/docs/reference-manual/index.html}}
}

@article{stephens:2000,
    author={Stephens, Matthew},
    year={2000},
    title={Dealing with label switching in mixture models}, 
    journal={Journal of the Royal
    Statistical Society, Series B},
    volume={62},
    pages={795-809}
}

@inproceedings{Akaike1973,
  author    = {Hirotugu Akaike},
  title     = {Information Theory and an Extension of the Maximum Likelihood Principle},
  booktitle = {Proceedings of the Second International Symposium on Information Theory},
  editor    = {B. N. Petrov and F. Cs{\'a}ki},
  pages     = {267--281},
  publisher = {Akad{\'e}miai Kiad{\'o}},
  address   = {Budapest},
  year      = {1973}
}

@book{Plummer2017,
	author = {Plummer, Martyn},
	title = {{JAGS Version 4.3.0 User Manual}},
	year = {2017},
        url = {https://sourceforge.net/projects/mcmc-jags/files/Manuals/4.x/jags_user_manual.pdf/download}
}

@book{OpenBUGS2010,
author = {Lambert M. Surhone and Mariam T. Tennoe and Susan F. Henssonow},
title = {{OpenBUGS}},
year = {2010},
isbn = {6133181206},
publisher = {Betascript Publishing},
address = {Beau Bassin, MUS}
}

@article{celeux:2006,
author = {Gilles Celeux and Florence Forbes and Christian P. Robert and Donald Michael Titterington},
title = {{Deviance information criteria for missing data models}},
volume = {1},
journal = {Bayesian Analysis},
number = {4},
pages = {651 -- 674},
year = {2006}
}

@article{Spiegelhalter2014,
    author = {Spiegelhalter, David J. and Best, Nicola G. and Carlin, Bradley P. and van der Linde, Angelika},
    title = {The Deviance Information Criterion: 12 Years on},
    journal = {Journal of the Royal Statistical Society Series B: Statistical Methodology},
    volume = {76},
    number = {3},
    pages = {485-493},
    year = {2014},
    month = {04},
    issn = {1369-7412},
    doi = {10.1111/rssb.12062},
    url = {https://doi.org/10.1111/rssb.12062},
}

@article{Xiao_Rabe-Hesketh_Skrondal_2025, 
title={Bayesian Identification and Estimation of Growth Mixture Models}, 
DOI={10.1017/psy.2025.11}, 
journal={Psychometrika}, 
author={Xiao, Xingyao and Rabe-Hesketh, Sophia and Skrondal, Anders}, 
year={2025}, 
pages={1–35}}

@article{gelman2014understanding,
  title={Understanding predictive information criteria for Bayesian models},
  author={Gelman, Andrew and Hwang, Jessica and Vehtari, Aki},
  journal={Statistics and Computing},
  volume={24},
  pages={997--1016},
  year={2014},
  publisher={Springer}
}

@article{rousseau:2011,
    author={J. Rousseau and K. Mengersen},
    year={2011},
    title={Asymptotic behaviour of the posterior distribution in overfitted mixture models},
    journal={Journal of the Royal Statistical Society, Series B},
    volume={73},
    pages={689-710}
}

@article{erosheva2017_FA,
  title={Dealing with reflection invariance in {Bayesian} factor analysis},
  author={Erosheva, Elena A and Curtis, S McKay},
  journal={Psychometrika},
  volume={82},
  number={2},
  pages={295--307},
  year={2017},
  publisher={Cambridge University Press \& Assessment}
}

@article{papastamoulis2022_FA,
  title={On the identifiability of {Bayesian} factor analytic models},
  author={Papastamoulis, Panagiotis and Ntzoufras, Ioannis},
  journal={Statistics and Computing},
  volume={32},
  number={2},
  pages={23},
  year={2022},
  publisher={Springer}
}

@book{mccullagh1987tensor,
  author    = {Peter McCullagh},
  title     = {Tensor Methods in Statistics},
  publisher = {Chapman \& Hall},
  address   = {London},
  year      = {1987}
}

@book{vanderVaart1998,
  author    = {Aad W. van der Vaart},
  title     = {Asymptotic Statistics},
  series    = {Cambridge Series in Statistical and Probabilistic Mathematics},
  publisher = {Cambridge University Press},
  address   = {Cambridge},
  year      = {1998}
}

@misc{muthen2010bayesian,
  title={Bayesian analysis in {Mplus}: A brief introduction},
  url={https://www.statmodel.com/download/IntroBayesVersion%203.pdf},
  author={Muth{\'e}n, Bengt},
  year={2010}
}

@article{Richardson:2002,
  author  = {Richardson, Sylvia},
  title   = {Discussion of ``{B}ayesian Measures of Model Complexity 
             and Fit'' by {S}piegelhalter, {B}est, {C}arlin, and 
             van der {L}inde},
  journal = {Journal of the Royal Statistical Society: Series B 
             (Statistical Methodology)},
  volume  = {64},
  number  = {4},
  pages   = {627},
  year    = {2002}
}

@manual{CmdStanManual2024,
  title  = {CmdStan User's Guide},
  author = {{Stan Development Team}},
  year   = {2024},
  url    = {https://mc-stan.org/docs/cmdstan-guide/}
}

@article{burkner2023posterior,
  title={posterior: Tools for working with posterior distributions},
  author={B{\"u}rkner, Paul-Christian and Gabry, Jonah and Kay, Matthew and Vehtari, Aki},
  journal={R package version},
  volume={1},
  number={0},
  year={2023}
}

@article{Du2024,
  author = {Du, Han and Keller, Brian and Alacam, Egamaria and Enders, Craig},
  title = {Comparing {DIC} and {WAIC} for multilevel models with missing data},
  journal = {Behavior Research Methods},
  volume = {56},
  pages = {2731--2750},
  year = {2024},
  doi = {10.3758/s13428-023-02231-0}
}

@article{Li2020,
  author = {Li, Yong and Zeng, Tao and Yu, Jun},
  title = {Deviance information criterion for latent variable models and misspecified models},
  journal = {Journal of Econometrics},
  volume = {216},
  number = {2},
  pages = {450--493},
  year = {2020}
}

\clearpage
\normalsize
\appendix

%==========================================================================
% APPENDIX: R IMPLEMENTATION
%==========================================================================
\section*{Appendix: R Function for Computing \DICpV}
\label{sec:r_code}

To facilitate the application of \DICpV, we provide a general-purpose R function below. The function takes a single argument, \texttt{log\_lik}, which is an $S \times N$ matrix of pointwise log-likelihood draws, where $S$ is the number of posterior draws (iterations) and $N$ is the number of observations. This matrix can be readily extracted from standard Bayesian software such as \texttt{rstan} (via \texttt{extract\_log\_lik}) or \texttt{cmdstanr} (via the \texttt{\$draws()} method). The function is also available in the \texttt{dicv} R package (\url{https://github.com/DoriaXiao/dicv}), which includes convenience wrappers for \texttt{cmdstanr} and a bundled Stan model for the factor analysis example.

\begin{figure}[H]
\begin{center}
\begin{minipage}{.95\linewidth}
\begin{myenv}{R Code: Variance-Based DIC}
\begin{lstlisting}
#' Compute Variance-Based DIC (DIC_pV)
#'
#' @param log_lik An S x N matrix of pointwise log-likelihoods.
#'   Rows (S) represent posterior MCMC draws.
#'   Columns (N) represent individual observations.
#' @return A list containing the DIC_V value, the penalty pV,
#'   and the posterior mean deviance E_D.

compute_dic_v <- function(log_lik) {

  # 1. Compute the marginal deviance for each MCMC draw (s)
  #    D(theta^s) = -2 * sum( log p(y_i | theta^s) )
  #    Row sums aggregate over N observations for each draw.
  deviance_draws <- -2 * rowSums(log_lik)

  # 2. Compute the posterior mean deviance (Goodness of Fit)
  #    E[D(theta)]
  E_D <- mean(deviance_draws)

  # 3. Compute the penalty: posterior variance of the deviance
  #    pV = 0.5 * Var(D(theta))
  pV <- 0.5 * var(deviance_draws)

  # 4. Compute DIC_pV
  dic_pv <- E_D + pV

  return(list(
    DIC_pV = dic_pv,
    pV     = pV,
    E_D    = E_D
  ))
}

# --- Example Usage ---
# library(cmdstanr)
# fit <- mod$sample(data = my_data, ...)
#
# # Extract S x N log-likelihood matrix
# log_lik_mat <- fit$draws("log_lik", format = "draws_matrix")
#
# # Compute indices
# results <- compute_dic_v(log_lik_mat)
# print(results$DIC_pV)
\end{lstlisting}
\end{myenv}
\end{minipage}
\end{center}
\end{figure}

\section*{Supplementary Materials}

\begin{figure}[h]
  \centering
  \includegraphics[width=\linewidth]{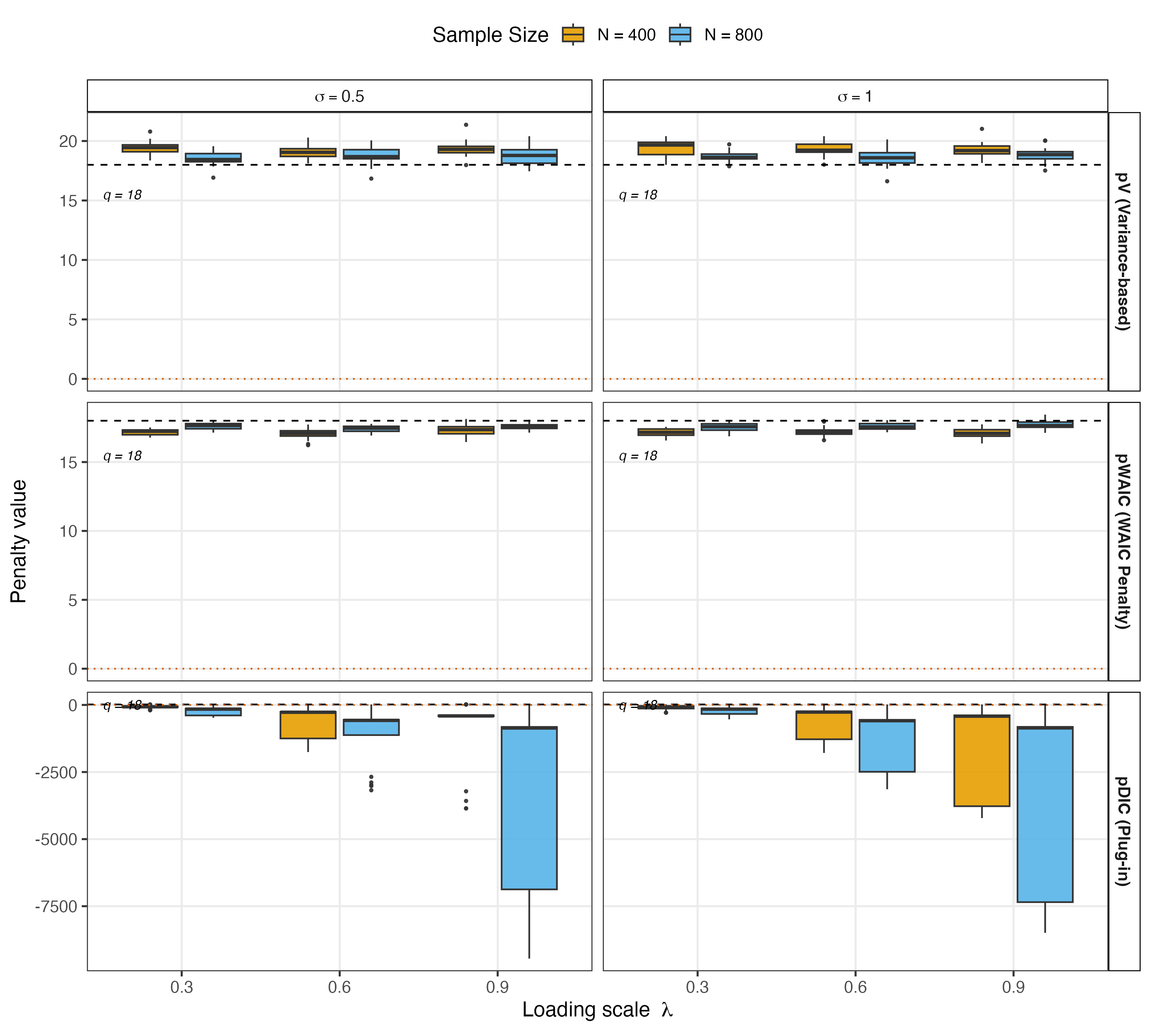}
  \caption{Full penalty stability figure with all 12
  conditions (6 panels: $\sigma \times$ penalty type).
  This is the expanded version of Figure~2 in the main
  text.}
  \label{fig:supp_penalty_full}
\end{figure}

\begin{figure}[h]
  \centering
  \includegraphics[width=\linewidth]{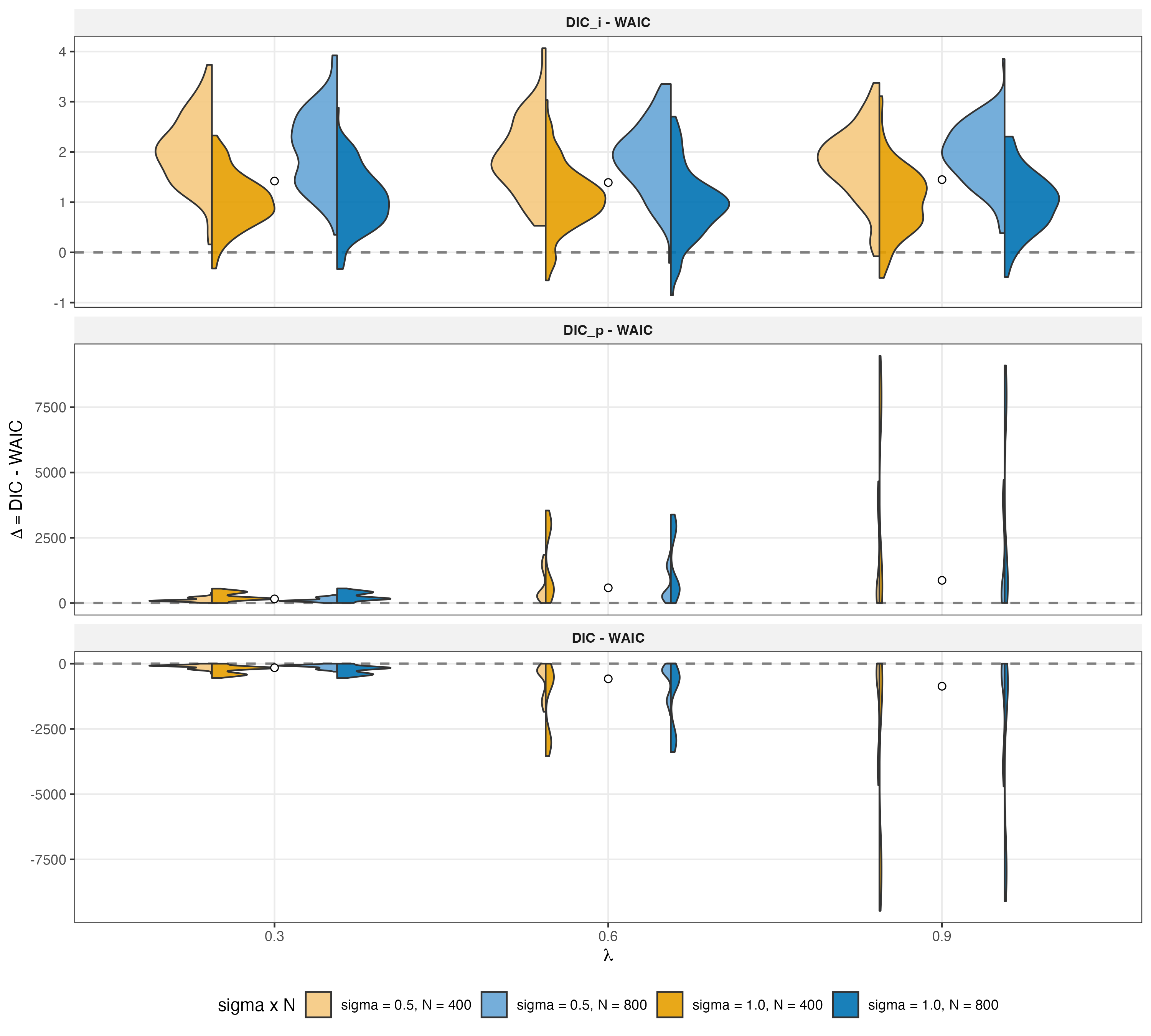}
  \caption{Distribution of differences between DIC
  variants and \WAIC\ ($\Delta = \text{Criterion} -
  \text{WAIC}$) across 1{,}200 factor analysis
  replications.}
  \label{fig:supp_delta_violin}
\end{figure}

\begin{figure}[h]
  \centering
  \includegraphics[width=\linewidth]{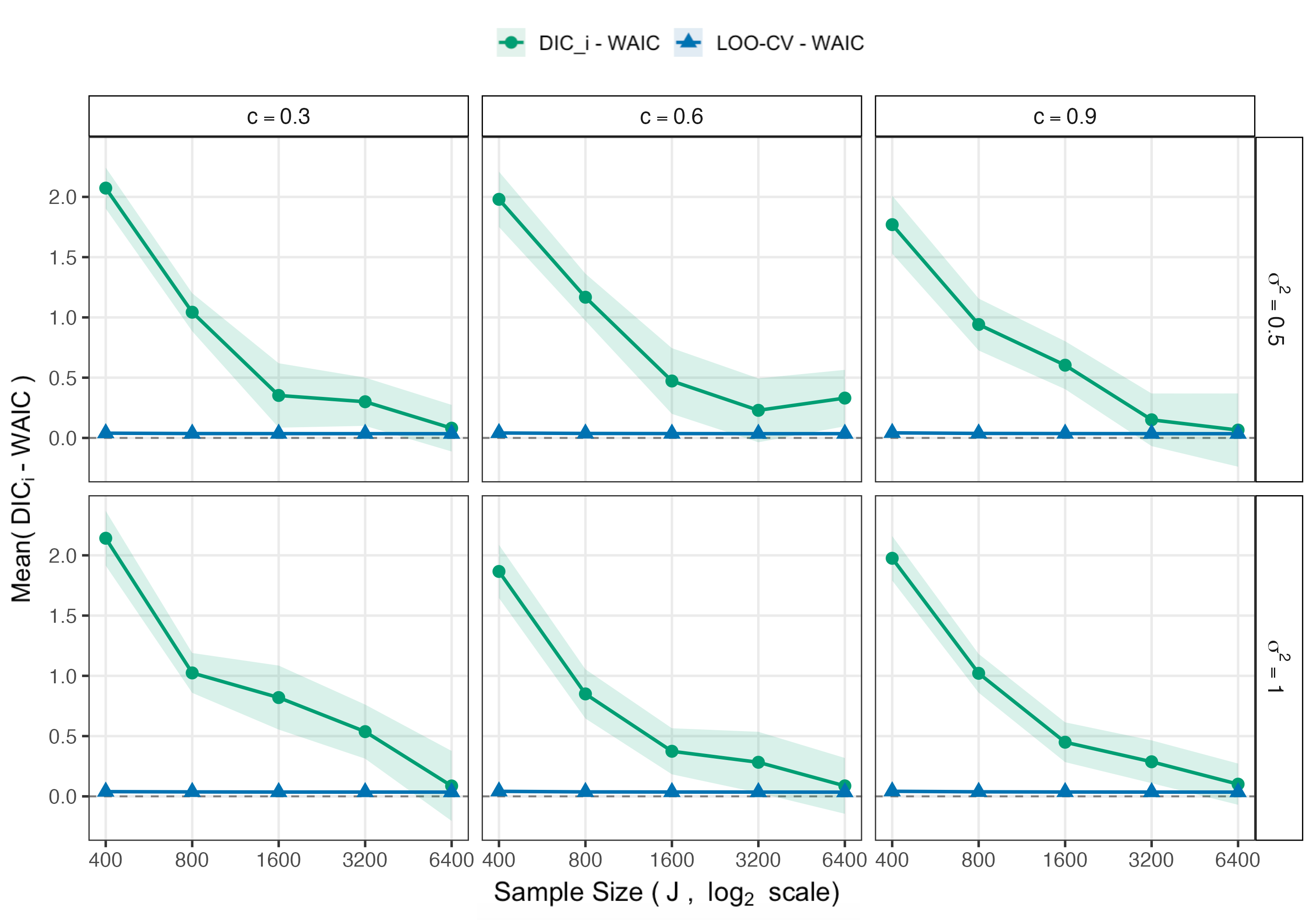}
  \caption{Mean difference $\DICpV - \WAIC$ across all
  six simulation conditions ($c \in \{0.3, 0.6, 0.9\}$,
  $\sigma^2 \in \{0.5, 1.0\}$). All conditions use 100
  replicates at $J = 400$ and $800$. At larger sample
  sizes ($J = 1{,}600$--$6{,}400$), the condition
  $c = 0.9$, $\sigma^2 = 1.0$ uses 100 replicates;
  the remaining five conditions use 20 replicates.
  This is the expanded version of
  Figure~\ref{fig:FA_convergence} in the main text.}
  \label{fig:supp_convergence_6panel}
\end{figure}

\begin{figure}[h]
  \centering
  \includegraphics[width=\linewidth]{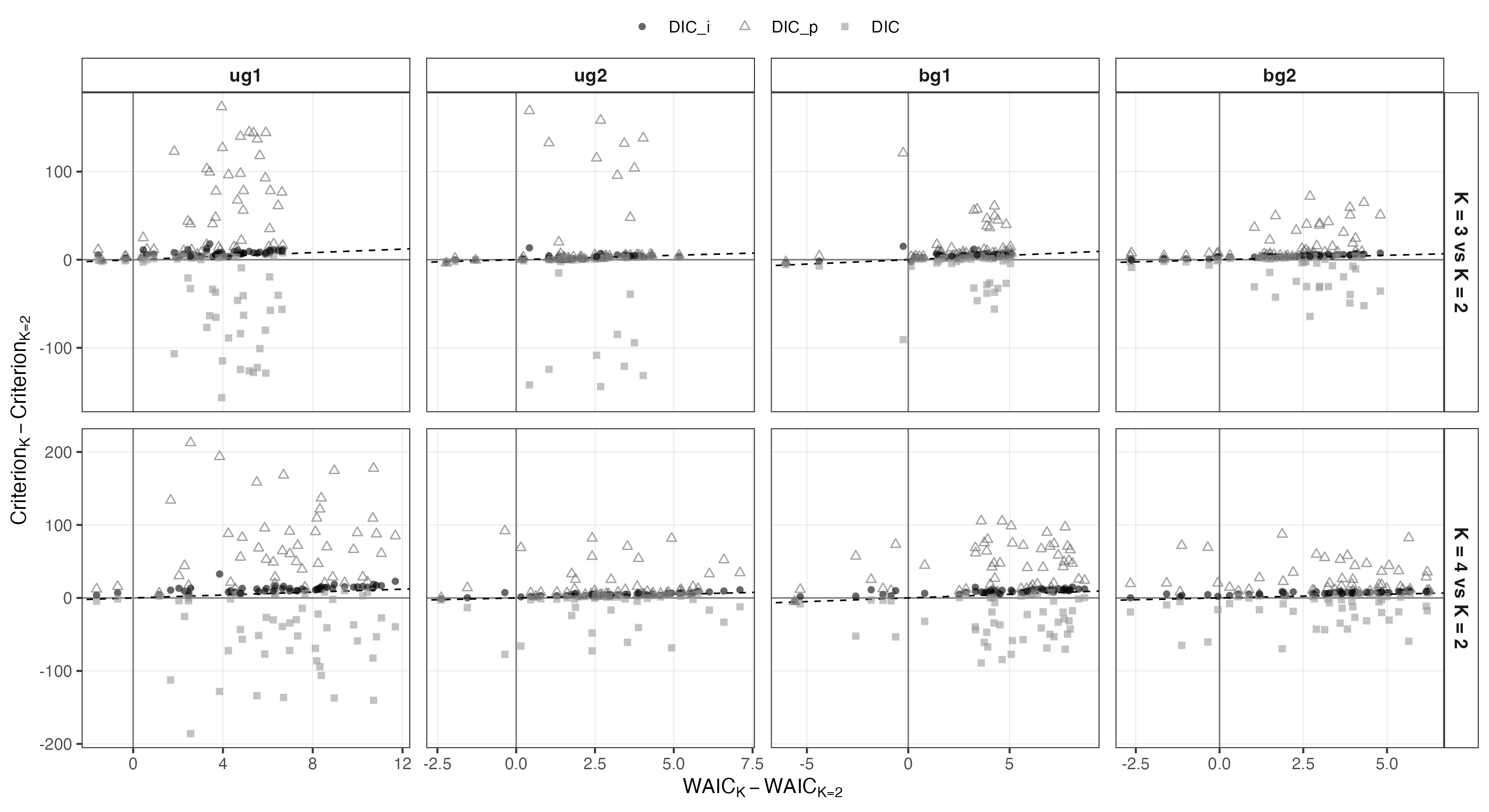}
  \caption{Differences in DIC variants versus differences in \WAIC\ for overfit candidate models relative to the true model ($K=2$), under the four greater-separation conditions. Top row: $K=3$ versus $K=2$. Bottom row: $K=4$ versus $K=2$. Three criteria are overlaid: \DICpV (filled circle), \DICpVtwo (open triangle), and classic DIC (light square). Each point represents one of 50 replicate datasets per condition. The dashed line is $y = x$. Solid gray lines mark the origin. The x-axis range is consistent within each column; the y-axis range is consistent within each row.}
  \label{fig:supp_3criteria_greater}
\end{figure}

\begin{figure}[h]
  \centering
  \includegraphics[width=\linewidth]{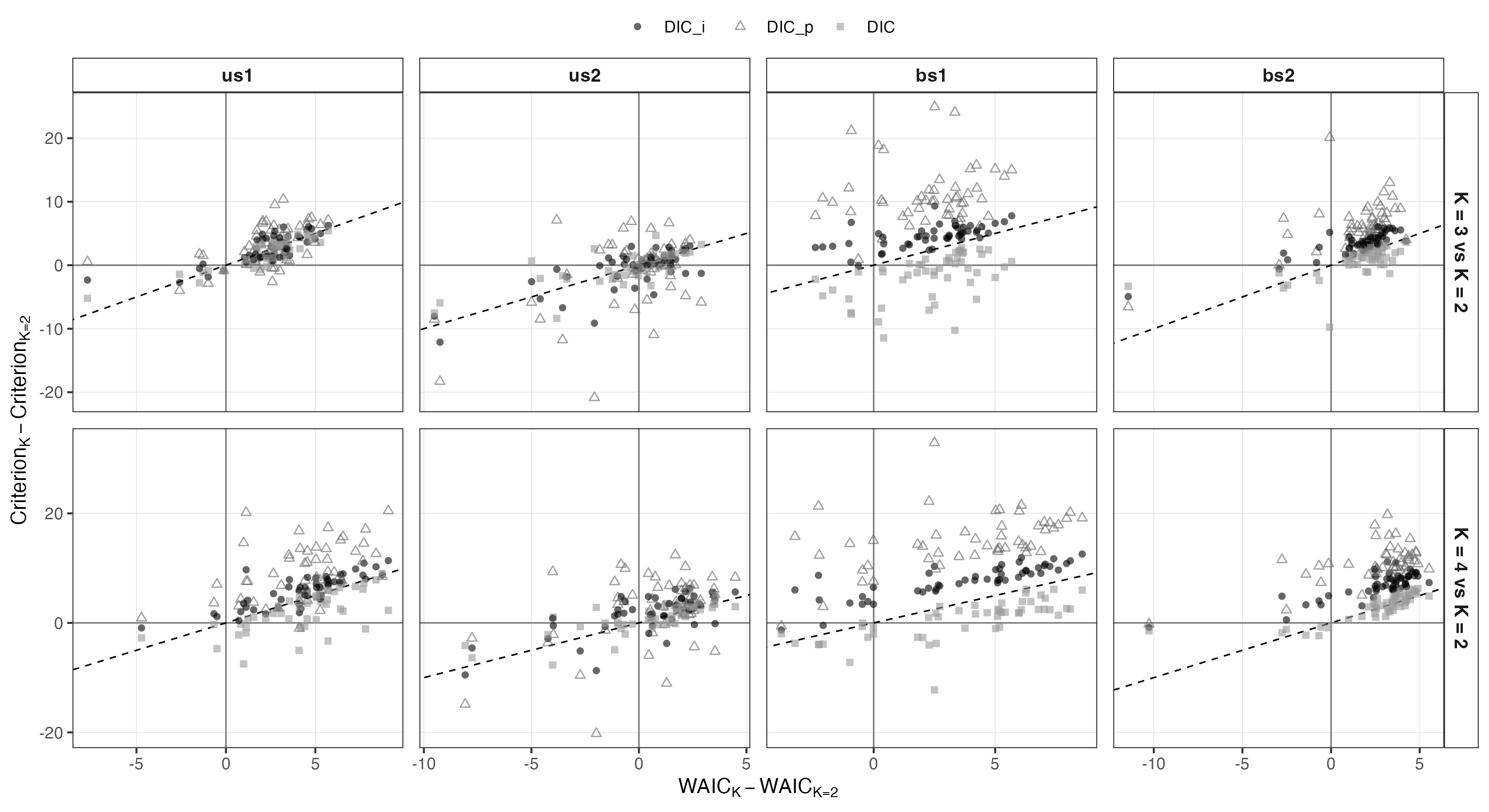}
  \caption{Differences in DIC variants versus differences in \WAIC\ for overfit candidate models relative to the true model ($K=2$), under the four smaller-separation conditions. Top row: $K=3$ versus $K=2$. Bottom row: $K=4$ versus $K=2$. Markers and reference lines as in Figure~\ref{fig:supp_3criteria_greater}.}
  \label{fig:supp_3criteria_smaller}
\end{figure}

\begin{figure}[h]
  \centering
  \includegraphics[width=\linewidth]{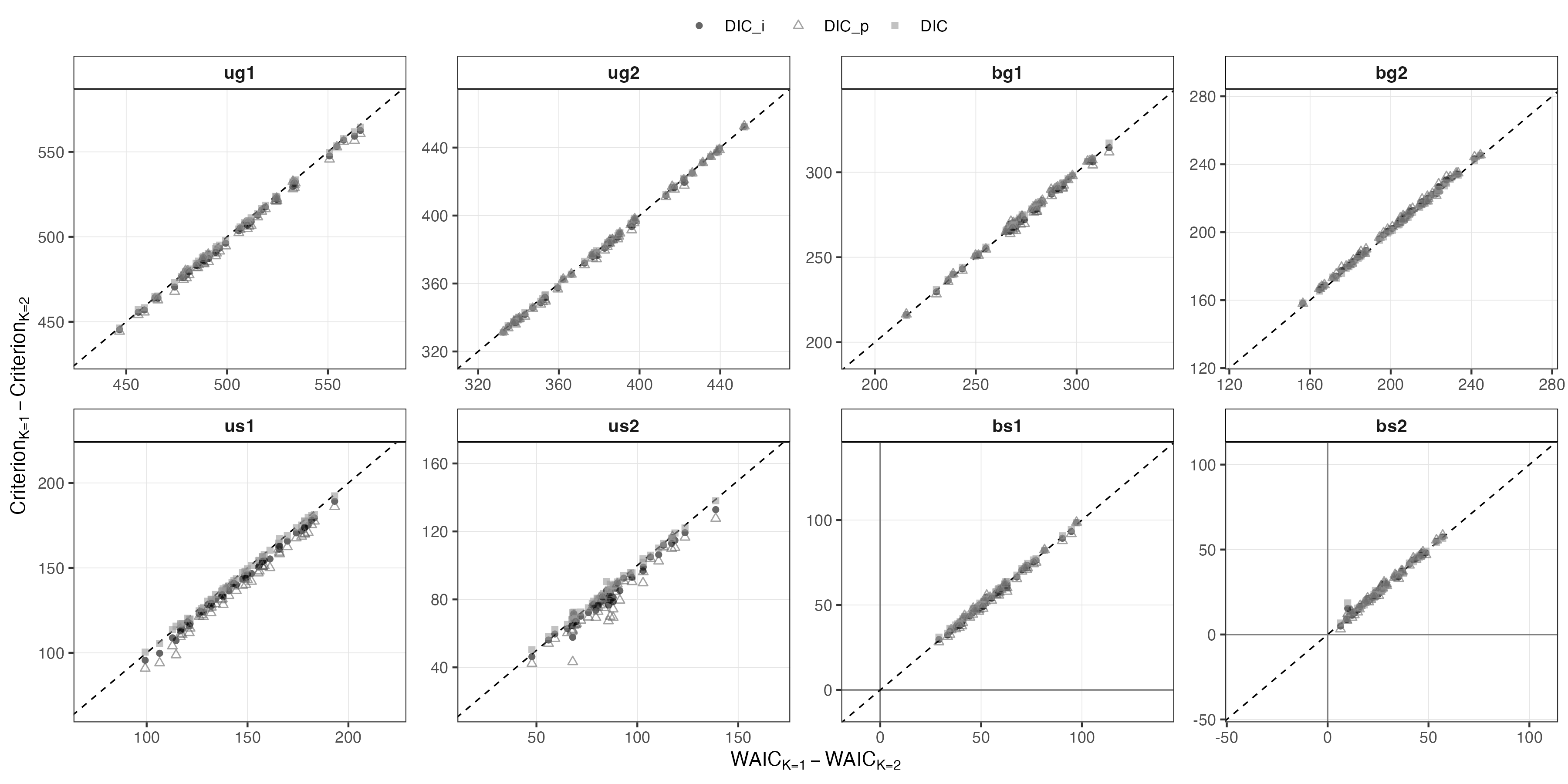}
  \caption{Differences in DIC variants versus differences in \WAIC\ for the underfit candidate model ($K=1$) relative to the true model ($K=2$), across all eight simulation conditions. Markers as in Figure~\ref{fig:supp_3criteria_greater}. Each panel uses a 150-unit window on both axes, centered on the panel's data; ranges are not required to include the origin.}
  \label{fig:supp_3criteria_K1}
\end{figure}

\end{document}